\documentclass[sigconf]{acmart}
\setcopyright{none}
\usepackage{listings}
\usepackage{booktabs}
\usepackage{array}

\usepackage[table]{xcolor}
\usepackage{tabularx}
\usepackage{adjustbox} 
\usepackage{supertabular}
\usepackage{longtable} 
\usepackage{caption}  

\newcommand{\pbg}[1]{\cellcolor{green!20}#1}
\newcommand{\fbg}[1]{\cellcolor{red!20}#1}

\AtBeginDocument{%
  }

\renewcommand\footnotetextcopyrightpermission[1]{}

\begin{document}

\title{Can Language Models Pass  Software Testing Certification Exams? a case study}

\author{Fitash Ul Haq}
\email{fitash.ulhaq@list.lu}
\affiliation{%
  \institution{Luxembourg Institute of Science and Technology (LIST)}
  \city{Esch-Sur-Alzette}
  \country{Luxembourg}
}

\author{Jordi Cabot}
\email{jordi.cabot@list.lu}
\affiliation{%
  \institution{Luxembourg Institute of Science and Technology (LIST)}
  \institution{University of Luxembourg}
  \city{Esch-Sur-Alzette}
  \country{Luxembourg}
}

\renewcommand{\shortauthors}{UL HAQ et al.}

\begin{abstract}
Large Language Models (LLMs) play a pivotal role in both academic research and broader societal applications. LLMs are increasingly used in software testing  activities such as test case generation, selection, and repair. However, several important questions remain: (1) do LLMs possess enough information about software testing principles to perform software testing tasks effectively? (2) do LLMs possess sufficient conceptual understanding of software testing to answer software testing questions under metamorphic transformations? and (3) do certain properties of software testing questions influence the performance of LLMs?

To answer these questions, this study evaluates 60 multimodal language models from both commercial vendors and the open-source community. 
The evaluation is performed using 30 sample exams of different types (core foundation, core advanced, specialist, and expert) from the International Software Testing Qualifications Board (ISTQB), which are used to assess the competence of human testers.  
In total, each model is evaluated on 1,171 questions.
Furthermore, to ensure sufficient conceptual understanding, the models are also tested on exam questions transformed using context-preserving metamorphic techniques.
Two models passed all the certifications by scoring at least 65\% in all of the 30 certification exams, with commercial models generally outperforming open-source ones. We analyze the reasons behind incorrect answers and provide recommendations for improving the design of software testing certification exams.


\end{abstract}



\maketitle

\section{Introduction}
Large Language Models (LLMs) have made significant advancements in recent years. They are now widely being adopted in multiple domains, including academic, industrial, and everyday applications. 
Prominent examples of LLM applications usage include chatbots, personal assistants, and machine translation systems. 
These models are trained on large scale datasets to perform tasks that involve understanding or generating natural language. 
Beyond natural language processing, modern language models are multimodal,  supporting the  processing and generation of multi-lingual text, images, videos and audios.

Pre-trained large language models are increasingly being applied in the domain of software testing to perform a variety of tasks, including test case generation \cite{schafer2023empirical,guilherme2023initial,zhang2023well,dakhel2024effective,tang2024chatGPT,hashtroudi2023automated}, test selection \cite{le2025testweaver,PanGhalebBriand2024_LTM,TaherkhaniHemmati2024_VALTEST}, oracle generation \cite{HossainDwyer2024_TOGLL,KhandakerKifetewPrandiSusi2025_AugmenTest,zhang2023algo,molina2025test} and test repair \cite{fatima2024flakyfix,yaraghi2025automated,liu2024fix,gu2024testart}.
Along with these practical applications, there is a growing interest in evaluating the extent to which these models possess an understanding of fundamental software testing principles, as measured by certification exams. 
More specifically, it is necessary to determine whether the language model possesses sufficient knowledge of software testing to perform software testing tasks effectively.
Furthermore, it is important to assess whether the model demonstrates a foundation, advanced, specialist, or expert level of understanding.
In addition, it is vital to evaluate the robust understanding of the software testing principles by examining how language models perform when certification exam questions undergo metamorphic transformations. 
This evaluation ensures that the model possesses conceptual understanding of software testing rather than the  memorization of specific question formulations.
Assessing their knowledge in this way provides insight into their reliability and potential for automating complex testing activities.

In this paper, we investigate the capabilities of language models in attempting the software testing certification exams. 
The rationale behind this investigation is the premise that such certification examinations serve as standardized instruments for assessing the competence of human software testers.
To systematically investigate the capabilities of LLMs, we examine them from different perspectives. 
First, we assess which language models can pass ISTQB certification exams. 
Second, we analyze how different models perform on questions across different learning objective levels. This perspective is important because it is possible that a model passes (i.e., achieves 65\% of the score) the exam by answering the higher-mark questions but performs poorly on other questions.
Third, we evaluate how models perform on the exam questions with metamorphic transformations to test robust conceptual understanding rather than memorization of LLMs. 
Finally, we investigate the properties of exam questions  that the models consistently answer incorrectly.

Specifically, we investigate 60 large language models of varying sizes from both commercial vendors and open-source community. 
We use a total of 30 International Software Testing Qualifications
Board (ISTQB) sample certification exams to evaluate these language models. This set of exams covers multiple levels, such as core foundation, core advanced, specialist, and expert.
We use certification exams from ISTQB because: (i) they are recognized globally, (ii) cover a wide range of software testing principles, (iii) the syllabus, sample exams, answers, exam structure, and rules are publicly available. 
The questions in exams have six levels, and marks are allocated based on the level. The higher the level, the more difficult the question. 
In total, for this study, every language model has to answer 1171 questions of different levels for the certification exams.

In summary, our contributions are as follows:
\begin{enumerate}
    \item A systematic comparison of 60 language models with regard to their capabilities in attempting to pass the certification exams.
    \item A systematic comparison of language models based on their ability to answer questions with different levels.
    \item An assessment of language model performance on exam questions with metamorphic transformations.
    \item A publicly available replication package to facilitate the reproducibility of our results. 
    \item An accessible and interactive public leaderboard showcasing LLM performance on certification exams~\footnote{\url{https://llm-certification-exam-dashboard-jclzzmhg68tdwkgk96yqzb.streamlit.app/}}.
\end{enumerate}

The rest of the paper is as follows: 
section~\ref{sec:background:ISTQB_certification_exam} presents a brief description of ISTQB certification exams.
Section~\ref{sec:methodology} presents an overview of methodology.
Section~\ref{sec:evaluation} presents the evaluation of language models in attempting the certification exams and further analyses their performance on questions with metamorphic transformations.
Section~\ref{threats} discusses the threats to validity of this study.
Section~\ref{sec:limitation} discusses the limitations of the language models observed during the experiments.
Section~\ref{sec:relatedworks} presents the literature review and shows how our work is different from the rest.
Finally, section \ref{sec:conclusion} concludes the paper and discusses the future work.

\section{ISTQB Certification Exams}\label{sec:background:ISTQB_certification_exam}
The International Software Testing Qualifications Board (ISTQB) is a non-profit, internationally recognized certification scheme for software testing professionals. 

ISTQB provides a standardized set of materials for software testing. 
ISTQB characterizes its certifications into four categories: Core Foundation, Core Advanced, Specialist, and  Expert. 
At the time of writing this paper,  ISTQB offers 29 certifications in total: 
one core foundation certification,
six core advanced certifications, 
17 Specialist certifications (subdivided into Technologies and Approaches with 9 certifications, Quality Characteristics \& Test Levels with 5 certifications, and Testing in Particular Domains with 3 certifications) and five expert certifications.
Table~\ref{table:ISTQBExam} shows a summary of the available certifications, their types, total questions, total points, and scoring points required to pass the exam. 
Note that for expert-level exams, we only consider multiple-choice questions, as open-ended essay questions are out of scope for this study.
To successfully obtain the certification, 65\% marks must be scored. Questions in the ISTQB exams are categorized into six levels based on their difficulty level~\cite{retzlaff2025automation}: 
\begin{enumerate}
    \item \textbf{K1:} Questions are designed to test the ability to recall or recognize a term.
    \item \textbf{K2:} Questions are designed to test the ability to understand or interpret a concept or topic.
    \item \textbf{K3:} Questions are designed to test the ability to apply learned concepts.
    \item \textbf{K4:} Questions are designed to test the ability to break down information for better understanding.
    \item \textbf{K5:} Questions are designed to test the ability to evaluate and make decisions based on defined criteria.
    \item \textbf{K6:} Questions are designed to test the ability to develop new solutions based on the acquired knowledge.
\end{enumerate}
ISTQB allows different time limits for different types of exams based on the types of questions included. 
The recommended allocation time depends  on the level of the question. 
Specifically, examinees are expected to spend approximately one minute for K1-level questions, one to two minutes for K2-level questions, three minutes for K3-level questions, and four minutes for K4-level questions.
K5 and K6 level questions are essay-based, and with a maximum of 90 minutes are recommended for such questions.
ISTQB allows an extra 25\% of time if the examinee is not a native english speaker.

\section{Methodology}\label{sec:methodology}
In this section, we provide an overview of our methodology of testing the capabilities of language models for attempting the certification exams.

\begin{figure}[h]
  \centering
  \includegraphics[width=\linewidth]{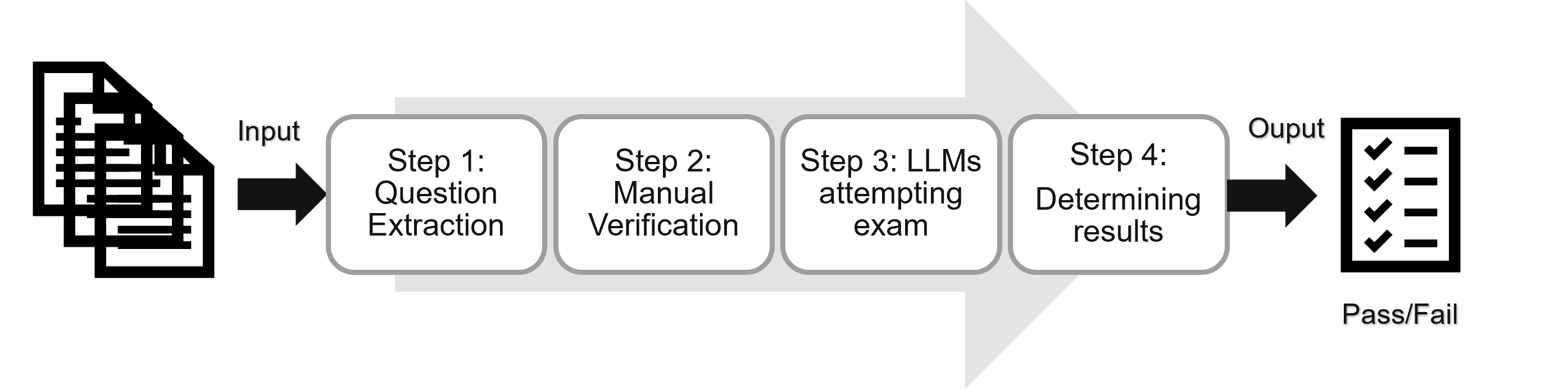}
  \caption{Overview of Methodology}
  \label{fig:methodology}
 \end{figure}

Figure~\ref{fig:methodology} shows the overview of the methodology used for this study. There are four steps: (1) Extracting questions, (2) Manual verification, (3) Querying the language model, and (4) Determining results.
In the first step, for each exam, the PDFs containing the questions and the answers are parsed to extract the following information: (i) questions, (ii) options,  (iii) number of possible answer options, (iv) marks for each question,  (v) level for each question, and the correct answer(s) for each question. 
In the second step, the extracted data is manually verified to ensure that the extracted data is correct. This step is important because the exam files do not have a consistent format. For example, some exams contain uppercase alphabetic enumeration and others lowercase alphabetic enumeration. In addition to an inconsistent format, some questions contain images that are extracted manually.
In the third step, a prompt is generated for each question to query the language model. We reused the prompt provided by Cedric et al.~\cite{lothritz2025testing} for language exam and adapted it for the ISTQB certification exams.
Listing~\ref{lst:prompt} shows an example of the prompt for the language model: it contains the instructions, a question, options, and the number of options to select. 
 
\lstset{
  breaklines=true,
  breakatwhitespace=true,
  postbreak=\mbox{\textcolor{red}{$\hookrightarrow$}\space},
}
\begin{lstlisting}[ caption={Prompt with example question for language model}, label={lst:prompt}]
I will give you a question from ISTQB exam, for each question I will specify how many options you can chose. You have to choose the best options. As output, simply write the chosen option, nothing else.
Question: Which of the following includes all the items that must be taken into account in a good test automation deployment strategy?
Options: 
a) Test environment; people; application access; test script storage; test data provisioning
... Other options omitted for brevity purposes
Select one option
\end{lstlisting}
In the final step, we compare the answers given by LLM with the ground truth to calculate whether the language model passes the test (i.e., scores more than or equal to 65\% or not).

\section{Evaluation}\label{sec:evaluation}
In this section, we present the evaluation of the capabilities of language models in attempting the ISTQB exams. 
Specifically, we answer the following three research questions:
\textbf{RQ1}: How do different language models fare in terms of passing the certification exams?
RQ1 aims to investigate the capabilities of different language models in attempting the certification exam. To do this, we first extract the questions and their corresponding answer options from the sample exams provided by the ISTQB. 
Then, we provide each question to different language models and record their response. 
Finally, we calculate the score for each language model and verify whether it has passed the exam (i.e., scored at least 65\%) or not.
\textbf{RQ2}: How do different language models fare in terms of attempting questions with different question levels?
RQ2 aims to investigate the capabilities of language models in solving questions with different levels of difficulty. 
The rationale behind this research question is that several models may perform similarly in terms of the number of exams passed.
This research question allows for a more detailed comparison between different language models.
As mentioned in section~\ref{sec:background:ISTQB_certification_exam}, questions can be of six levels, K1 to K6. In this study, we focus on levels K1 to K4, as K5 and K6 are open-ended essay questions, which are out of scope for this study.
To answer this RQ, we analyze the performance of the language model across different question levels and report which language model is best for each level. 
For RQ1 and RQ2, the questions are from sample exams that are available online, and
it may be possible that some LLMs have used these documents as part of their training dataset. 
To tackle this problem, we introduce RQ3, which transforms the questions without changing the underlying meaning of the question to verify whether LLM really understands the software testing principles or it merely memorises the answers available online. 
\textbf{RQ3}: How do best performing language models fare when ISTQB exam questions undergo metamorphic transformations, reflecting robust understanding rather than memorization?
RQ3 aims to investigate the performance of models when exam questions undergo metamorphic transformations. Specifically, we apply semantics-preserving transformations (e.g., synonym substitution and option reordering) to a set of selected questions from ISTQB exams. We then compare the performance of models on original and transformed questions to assess whether models rely on robust conceptual understanding of software testing principles or memorization of specific question formulations.

\subsection{Experimental Subjects}
We conduct our experiments on a large number of multimodal language models.
Multimodal models are used because some of the exam questions include visual components such as images and tables.
In total, we evaluated on 60 large language models of different sizes from different commercial vendors and open-source community. Table~\ref{table:RQ1:result-1} and ~\ref{table:RQ1:result-2}  provides a complete list of the  commercial and open-source models included in the study.
Note that some models include a timestamp in their name; this timestamp shows the date on which that model was released. For example, GPT-5 and GPT-5-08-07 share the same underlying model architecture and parameters but are released on different dates with minor changes.
GPT-5 is the latest version with the updates to better recognize and support people in moments of distress~\cite{openaiChangenotes} while GPT-5-08-07 is a snapshot of the model released on August 7, 2025.
These snapshots are important for reproducibility as performance can vary across model releases due to such updates. 
In particular, our evaluation covers models from three commercial providers, OpenAI, Anthropic and Google (Gemini), in addition to multiple open-source alternatives. 
The OpenAI models include: GPT-5.2~\cite{openai2025gpt52}, GPT-5.1~\cite{openai2025gpt51}, GPT-5 and its variants~\cite{openai2025gpt5,openai2025gpt5mini,openai2025gpt5nano},
GPT-4.1 and its variants~\cite{openai2025gpt41,openai2025gpt41mini,openai2025gpt41nano}, GPT-4 and its variants~\cite{openai2025gpt4,openai2025gpt4turbo}, O4 mini~\cite{openai2025o4}, O3~\cite{openai2025o3}, GPT-4O and its variants~\cite{openai2025o4o}, ChatGPT-4O-latest~\cite{openai20254o}.
The Gemini models include Gemini-3-flash-preview~\cite{Gemini30}, Gemini-2.5-pro~\cite{comanici2025gemini}, Gemini-2.5-flash~\cite{comanici2025gemini}, Gemini-2.5-flash-lite~\cite{comanici2025gemini}, Gemini-2.5-flash-lite-preview~\cite{comanici2025gemini}, Gemini-2.0-flash~\cite{comanici2025gemini}, and Gemini-2.0-flash-lite~\cite{comanici2025gemini}.
From the Anthropic, we evaluated the Claude Sonnet 4.5, Claude-3-Haiku, Claude-4.5-Haiku and Claude-Opus-4.5.  
Finally, the open source models include: LLama 4 Maverick, Nemotron Nano 2~\cite{basant2025nvidia} , Ministral 3 and its variants~\cite{liu2026ministral}, LLaVA and its variants ~\cite{liu2024improved}, Qwen 2.5 and its variants~\cite{bai2025qwen2}, Gemma3 and its variants~\cite{team2025gemma},
Mistral-small3.1~\cite{mistralsmall3.1}, Mistral-small3.2~\cite{mistralsmall3.2}, granite3.2-vision:2b~\cite{team2025granite}, minicpm-v~\cite{yao2024minicpm},llava-llama3~\cite{2023xtuner}, llava-phi3~\cite{2023xtuner} and bakllava~\cite{2023xtuner}.
We initially included  Gemini-3-pro, Qwen3-VL, and its variants in our evaluation. 
However, Gemini-3-pro has lower request per day limit, which prevented the completion of the evaluation within due time. Consequently, it was excluded from the set of evaluated LLMs, but Gemini-3-pro will be included in the future studies.
Qwen3-VL and its variants were taking too long to respond to the prompt and were ending in request timeout. As a result, these models were also excluded  from our evaluation.

\begin{table}[h!]
\centering
\caption{Distribution of K1–K4 Questions per Certification Exam}
\label{table:klevel-counts}
\renewcommand{\arraystretch}{1.1}
\setlength{\tabcolsep}{4pt}
\scriptsize

\begin{minipage}[t]{0.48\columnwidth}
\centering
\begin{tabular}{lcccc}
\toprule
\textbf{Test} & \textbf{K1} & \textbf{K2} & \textbf{K3} & \textbf{K4} \\
\midrule
CTFLvA & 8 & 24 & 8 & 0 \\
CTFLvB & 8 & 24 & 8 & 0 \\
CTFLvC & 8 & 24 & 8 & 0 \\
CTFLvD & 8 & 24 & 8 & 0 \\

CTAL-TAv4.1 & 0 & 18 & 21 & 6 \\
CTAL-TAv3.1 & 0 & 17 & 6 & 17 \\
CTAL-TAEv2.0 & 0 & 20 & 15 & 5 \\
CTAL-TAEv1.3 & 0 & 17 & 11 & 12 \\
CTAL-TMv3.0 & 0 & 26 & 10 & 14 \\
CTAL-TMv1.4 & 0 & 32 & 14 & 10 \\
CTAL-ATT & 0 & 13 & 4 & 3 \\
CTAL-TTA & 0 & 20 & 17 & 8 \\
CT-ATLaS & 0 & 20 & 9 & 11 \\
CT-AI & 4 & 29 & 3 & 4 \\
CTFL-AT & 10 & 22 & 8 & 0 \\

\bottomrule
\end{tabular}
\end{minipage}%
\hfill
\begin{minipage}[t]{0.48\columnwidth}
\centering
\begin{tabular}{lcccc}
\toprule
\textbf{Test} & \textbf{K1} & \textbf{K2} & \textbf{K3} & \textbf{K4} \\
\midrule
CT-GenAI & 8 & 26 & 6 & 0 \\
CT-MAT & 11 & 23 & 6 & 0 \\

CT-TAS & 0 & 31 & 9 & 0 \\
CT-MBT & 13 & 22 & 5 & 0 \\
CT-AcT & 3 & 27 & 10 & 0 \\
CT-PT & 2 & 25 & 2 & 11 \\
CT-SEC & 0 & 19 & 16 & 10 \\
CT-STE & 0 & 24 & 13 & 3 \\
CT-UT & 6 & 28 & 3 & 3 \\
CT-AuT & 14 & 19 & 7 & 0 \\
CT-GaMe & 6 & 30 & 4 & 0 \\
CT-GT & 5 & 35 & 0 & 0 \\
CTEL-ITP-ATP & 0 & 16 & 7 & 2 \\
CTEL-ITP-ITPI & 0 & 17 & 3 & 5 \\
CTEL-TM-OTM & 0 & 5 & 4 & 11 \\
\bottomrule
\end{tabular}
\end{minipage}
\end{table}

\begin{table*}[t]

\caption{Results of Commercial Language models on different ISTQB Exams. On x-axis are the certification exams, on y-axis are the models; each cell shows the percentage scored by the model}
\label{tab:models-istqb-1}
\renewcommand{\arraystretch}{1.15}
\setlength{\tabcolsep}{1.95pt}
\scriptsize
\centering
\renewcommand{\arraystretch}{1.1}
\rowcolors{2}{white}{gray!10}
\begin{tabular}{p{2cm}cccccccccccccccccccccccccccccccc}
\toprule
\textbf{Model} &
\rotatebox{90}{CTFL - A} &
\rotatebox{90}{CTFL - B} &
\rotatebox{90}{CTFL - C} &
\rotatebox{90}{CTFL - D} &
\rotatebox{90}{CTAL-TAv4.1} &
\rotatebox{90}{CTAL-TAv3.1} &
\rotatebox{90}{CTAL-TAEv2.0} &
\rotatebox{90}{CTAL-TAEv1.3} &
\rotatebox{90}{CTAL-TMv3.0} &
\rotatebox{90}{CTAL-TMv1.4} &
\rotatebox{90}{CTAL-ATT} &
\rotatebox{90}{CTAL-TTA} &
\rotatebox{90}{CT-ATLaS} &
\rotatebox{90}{CT-AI} &
\rotatebox{90}{CTFL-AT} &

\rotatebox{90}{CT-GenAI} &
\rotatebox{90}{CT-MAT} &
\rotatebox{90}{CT-TAS} &
\rotatebox{90}{CT-MBT} &

\rotatebox{90}{CT-AcT} &
\rotatebox{90}{CT-PT} &

\rotatebox{90}{CT-SEC} &
\rotatebox{90}{CT-STE} &
\rotatebox{90}{CT-UT} &
\rotatebox{90}{CT-AuT} &
\rotatebox{90}{CT-GaMe} &

\rotatebox{90}{CT-GT} &
\rotatebox{90}{CTEL-ITP-ATP} &
\rotatebox{90}{CTEL-ITP-ITPI} &
\rotatebox{90}{CTEL-TM-OTM} & \rotatebox{90}{Total Passed} 

\\
\midrule
\multicolumn{31}{l}{\textit{\textbf{OpenAI Models}}}\\
gpt-5.2&\pbg{88.8}&\pbg{73.8}&\pbg{80}&\pbg{83.8}&\fbg{64.1}&\pbg{75.6}&\pbg{80.4}&\pbg{84}&\pbg{86.4}&\pbg{83.5}&\pbg{71.7}&\pbg{76.3}&\pbg{87}&\pbg{85.1}&\pbg{87.5}&\pbg{95.7}&\pbg{82.5}&\pbg{76.5}&\pbg{87.5}&\pbg{72.5}&\pbg{80}&\pbg{77.5}&\pbg{82.6}&\pbg{85}&\pbg{85}&\fbg{62.5}&\fbg{60}&\pbg{69.4}&\pbg{86.8}&\pbg{74.6}&27
\\
\hline
gpt-5.1&\pbg{82.5}&\pbg{68.8}&\pbg{80}&\pbg{77.5}&\fbg{56.4}&\pbg{65}&\pbg{78.3}&\pbg{85.3}&\pbg{77.3}&\pbg{83.5}&\pbg{66.7}&\pbg{76.3}&\pbg{85.5}&\pbg{80.9}&\pbg{85}&\pbg{93.5}&\pbg{85}&\pbg{82.7}&\pbg{88.8}&\pbg{85}&\pbg{77.5}&\pbg{85}&\pbg{74.4}&\pbg{90}&\pbg{92.5}&\pbg{70}&\fbg{62.5}&\pbg{72.2}&\pbg{89.5}&\pbg{83.3}&28
\\
\hline
GPT-5&\pbg{94.6}&\pbg{87.5}&\pbg{91.2}&\pbg{91.2}&\pbg{95.1}&\pbg{83.5}&\pbg{82.9}&\pbg{86.9}&\pbg{79.2}&\pbg{83.7}&\pbg{76.1}&\pbg{89.3}&\pbg{83.1}&\pbg{88.7}&\pbg{89.6}&\pbg{94.9}&\pbg{88.3}&\pbg{79.6}&\pbg{98.3}&\pbg{87.5}&\pbg{75.8}&\pbg{77.9}&\pbg{80.2}&\pbg{88.3}&\pbg{93.3}&\pbg{68.8}&\fbg{64.2}&\pbg{80.6}&\pbg{93.9}&\pbg{91.8}&29
\\
\hline
gpt-5-mini&\pbg{90}&\pbg{88.3}&\pbg{83.8}&\pbg{90}&\pbg{84}&\pbg{72.5}&\pbg{74.6}&\pbg{80}&\pbg{76.9}&\pbg{80.4}&\pbg{73.3}&\pbg{85}&\pbg{89.6}&\pbg{80.1}&\pbg{88.3}&\pbg{92.8}&\pbg{90}&\pbg{84.4}&\pbg{91.7}&\pbg{85}&\pbg{75.8}&\pbg{80.8}&\pbg{71.7}&\pbg{84.2}&\pbg{87.5}&\pbg{68.3}&\fbg{62.5}&\pbg{79.6}&\pbg{88.6}&\pbg{83}&29
\\
\hline
gpt-5-nano&\pbg{79.2}&\pbg{73.8}&\pbg{74.6}&\pbg{86.2}&\pbg{70.1}&\fbg{55.4}&\fbg{62.3}&\pbg{79.1}&\pbg{67}&\pbg{67.1}&\pbg{70.6}&\pbg{74.4}&\pbg{75.4}&\pbg{77.3}&\pbg{81.7}&\pbg{66}&\pbg{70}&\pbg{69}&\pbg{81.7}&\pbg{81.7}&\pbg{67.5}&\pbg{66.7}&\pbg{68.6}&\pbg{75}&\pbg{75.8}&\fbg{62.9}&\fbg{57.5}&\pbg{74.1}&\fbg{64.9}&\fbg{53.6}&24
\\
\hline
gpt-5-2025-08-07&\pbg{93.8}&\pbg{87.1}&\pbg{90}&\pbg{91.2}&\pbg{95.1}&\pbg{82.3}&\pbg{78.5}&\pbg{86.7}&\pbg{80.7}&\pbg{82.7}&\pbg{76.7}&\pbg{87.2}&\pbg{85.7}&\pbg{87.2}&\pbg{93.8}&\pbg{71.7}&\pbg{85.8}&\pbg{82.3}&\pbg{98.8}&\pbg{86.7}&\pbg{77.5}&\pbg{78.3}&\pbg{77.9}&\pbg{89.2}&\pbg{93.3}&\pbg{66.2}&\pbg{67.5}&\pbg{78.7}&\pbg{93.9}&\pbg{85.1}&30
\\
\hline
gpt-5-mini-2025-08-07&\pbg{89.2}&\pbg{90.4}&\pbg{84.6}&\pbg{90.4}&\pbg{82.1}&\pbg{76.9}&\pbg{74.9}&\pbg{77.8}&\pbg{78.8}&\pbg{80}&\pbg{72.2}&\pbg{87.2}&\pbg{85.7}&\pbg{80.9}&\pbg{89.6}&\pbg{66.8}&\pbg{91.2}&\pbg{81.3}&\pbg{92.9}&\pbg{86.7}&\pbg{75}&\pbg{73.8}&\pbg{77.9}&\pbg{82.5}&\pbg{86.7}&\pbg{65}&\fbg{55.8}&\pbg{83.3}&\pbg{86}&\pbg{76.2}&29
\\
\hline
gpt-5-nano-2025-08-07&\pbg{85.4}&\pbg{73.3}&\pbg{72.9}&\pbg{82.1}&\pbg{68.4}&\fbg{50.2}&\fbg{58.9}&\pbg{78.7}&\fbg{64.6}&\pbg{66.5}&\pbg{73.3}&\pbg{76.7}&\pbg{74.4}&\pbg{73}&\pbg{79.6}&\pbg{66.6}&\pbg{73.8}&\pbg{68}&\pbg{80.8}&\pbg{82.5}&\pbg{71.7}&\pbg{69.2}&\pbg{70.2}&\pbg{72.5}&\pbg{80}&\pbg{66.2}&\fbg{58.3}&\pbg{74.1}&\pbg{70.2}&\fbg{50.9}&25
\\
\hline
gpt-5-chat-latest&\pbg{87.5}&\pbg{72.5}&\pbg{82.5}&\pbg{81.2}&\pbg{70.5}&\fbg{61.3}&\pbg{72.5}&\pbg{88}&\pbg{73.9}&\pbg{86.5}&\pbg{78.3}&\pbg{83.3}&\pbg{92.8}&\pbg{76.6}&\pbg{87.5}&\pbg{93.5}&\pbg{82.5}&\pbg{75.5}&\pbg{90}&\pbg{85}&\pbg{75}&\pbg{80}&\pbg{80.2}&\pbg{87.5}&\pbg{90}&\pbg{70}&\fbg{62.5}&\pbg{69.4}&\pbg{81.6}&\pbg{74.6}&28
\\
\hline
gpt-4.1&\pbg{86.2}&\pbg{73.8}&\pbg{72.5}&\pbg{81.2}&\fbg{60.3}&\pbg{68.1}&\pbg{76.8}&\pbg{82.7}&\pbg{76.1}&\pbg{84.7}&\pbg{76.7}&\pbg{74.4}&\pbg{92.8}&\pbg{80.9}&\pbg{86.2}&\pbg{95.7}&\pbg{80}&\pbg{72.4}&\pbg{95}&\pbg{82.5}&\pbg{75}&\pbg{68.8}&\pbg{75.6}&\pbg{87.5}&\pbg{87.5}&\pbg{70}&\pbg{65}&\pbg{69.4}&\pbg{86.8}&\pbg{76.7}&29
\\
\hline
gpt-4.1-mini&\pbg{75}&\pbg{65}&\pbg{77.5}&\pbg{70}&\fbg{62.8}&\fbg{55}&\pbg{76.8}&\pbg{81.3}&\pbg{71}&\pbg{81.8}&\pbg{68.3}&\pbg{75.6}&\pbg{91.3}&\pbg{76.6}&\pbg{80}&\pbg{89.1}&\pbg{75}&\pbg{77.6}&\pbg{91.2}&\pbg{77.5}&\pbg{75}&\pbg{71.2}&\pbg{79.1}&\pbg{80}&\pbg{87.5}&\pbg{66.2}&\fbg{62.5}&\pbg{72.2}&\pbg{81.6}&\pbg{73}&27
\\
\hline
gpt-4.1-nano&\pbg{67.5}&\fbg{60}&\fbg{60}&\pbg{70}&\fbg{39.7}&\fbg{51.2}&\pbg{68.1}&\pbg{68.7}&\pbg{67.6}&\pbg{71.2}&\pbg{68.3}&\fbg{50.6}&\pbg{81.9}&\pbg{68.1}&\pbg{75}&\pbg{81.5}&\pbg{77.5}&\fbg{60.2}&\pbg{70}&\pbg{72.5}&\fbg{62.5}&\fbg{61.3}&\pbg{69.8}&\pbg{67.5}&\pbg{67.5}&\pbg{68.8}&\fbg{55}&\fbg{58.3}&\fbg{52.6}&\fbg{43.7}&18
\\
\hline
gpt-4.1-2025-04-14&\pbg{83.8}&\pbg{73.8}&\pbg{75}&\pbg{83.8}&\fbg{60.3}&\pbg{68.1}&\pbg{76.8}&\pbg{82.7}&\pbg{76.1}&\pbg{85.9}&\pbg{66.7}&\pbg{74.4}&\pbg{91.3}&\pbg{80.9}&\pbg{83.8}&\pbg{95.7}&\pbg{80}&\pbg{74.5}&\pbg{92.5}&\pbg{82.5}&\pbg{77.5}&\pbg{72.5}&\pbg{73.3}&\pbg{87.5}&\pbg{87.5}&\pbg{70}&\pbg{65}&\pbg{69.4}&\pbg{89.5}&\pbg{83.3}&29
\\
\hline
gpt-4.1-mini-2025-04-14&\pbg{77.5}&\fbg{63.7}&\pbg{75}&\pbg{72.5}&\pbg{65.4}&\fbg{53.1}&\pbg{76.8}&\pbg{81.3}&\pbg{71}&\pbg{80.6}&\pbg{78.3}&\pbg{73.1}&\pbg{91.3}&\pbg{74.5}&\pbg{80}&\pbg{91.3}&\pbg{75}&\pbg{75.5}&\pbg{88.8}&\pbg{72.5}&\pbg{75}&\pbg{71.2}&\pbg{81.4}&\pbg{80}&\pbg{90}&\pbg{66.2}&\fbg{62.5}&\pbg{72.2}&\pbg{81.6}&\pbg{73}&27
\\
\hline
gpt-4.1-nano-2025-04-14&\pbg{67.5}&\fbg{57.5}&\fbg{60}&\pbg{70}&\fbg{42.3}&\fbg{47.5}&\pbg{71}&\pbg{68.7}&\pbg{66.5}&\pbg{68.2}&\fbg{61.7}&\fbg{53.2}&\pbg{81.9}&\pbg{68.1}&\pbg{72.5}&\pbg{81.5}&\pbg{80}&\fbg{62.2}&\pbg{70}&\pbg{75}&\fbg{62.5}&\fbg{62.5}&\fbg{64}&\pbg{67.5}&\pbg{70}&\pbg{66.2}&\fbg{55}&\fbg{58.3}&\fbg{42.1}&\fbg{41.5}&16
\\
\hline
o4-mini&\pbg{90.8}&\pbg{85.4}&\pbg{83.3}&\pbg{88.3}&\pbg{83.8}&\pbg{76}&\pbg{72.7}&\pbg{78.2}&\pbg{76.5}&\pbg{82.4}&\pbg{67.8}&\pbg{85}&\pbg{87}&\pbg{85.1}&\pbg{83.8}&\pbg{94.6}&\pbg{85.8}&\pbg{77.2}&\pbg{95}&\pbg{82.5}&\pbg{77.5}&\pbg{82.1}&\pbg{76.7}&\pbg{84.2}&\pbg{88.3}&\pbg{67.1}&\fbg{59.2}&\pbg{76.9}&\pbg{77.2}&\pbg{82.5}&29
\\
\hline
o4-mini-2025-04-16&\pbg{90.8}&\pbg{87.9}&\pbg{84.6}&\pbg{87.1}&\pbg{78.8}&\pbg{79.8}&\pbg{74.6}&\pbg{75.6}&\pbg{74.2}&\pbg{81.4}&\pbg{66.7}&\pbg{87.8}&\pbg{85.7}&\pbg{84.4}&\pbg{84.2}&\pbg{94.9}&\pbg{85.8}&\pbg{78.6}&\pbg{95.8}&\pbg{84.2}&\pbg{79.2}&\pbg{77.1}&\pbg{76.7}&\pbg{83.3}&\pbg{87.5}&\fbg{64.2}&\fbg{56.7}&\pbg{77.8}&\pbg{86.8}&\pbg{84.7}&28
\\
\hline
o3&\pbg{92.9}&\pbg{91.2}&\pbg{84.2}&\pbg{89.2}&\pbg{87.2}&\pbg{79.6}&\pbg{73.7}&\pbg{90.2}&\pbg{74.2}&\pbg{79}&\pbg{76.7}&\pbg{85.7}&\pbg{75.4}&\pbg{78.7}&\pbg{91.2}&\pbg{94.2}&\pbg{85}&\pbg{78.6}&\pbg{94.6}&\pbg{88.3}&\pbg{76.7}&\pbg{75}&\pbg{67.1}&\pbg{85.8}&\pbg{93.3}&\pbg{72.1}&\fbg{60}&\pbg{80.6}&\pbg{87.7}&\pbg{71.2}&29
\\
\hline
o3-2025-04-16&\pbg{90}&\pbg{88.8}&\pbg{87.1}&\pbg{90}&\pbg{86.1}&\pbg{81.5}&\pbg{73.2}&\pbg{82}&\pbg{75.2}&\pbg{76.5}&\pbg{77.2}&\pbg{85.5}&\pbg{75.6}&\pbg{80.1}&\pbg{88.3}&\pbg{94.9}&\pbg{86.7}&\pbg{80.3}&\pbg{94.6}&\pbg{86.7}&\pbg{77.5}&\pbg{75}&\pbg{69.4}&\pbg{87.5}&\pbg{92.5}&\pbg{67.5}&\fbg{64.2}&\pbg{75.9}&\pbg{90.4}&\pbg{77.4}&29
\\
\hline
gpt-4o&\pbg{83.8}&\pbg{71.2}&\pbg{72.5}&\pbg{67.5}&\fbg{52.6}&\fbg{46.9}&\pbg{76.1}&\pbg{80}&\pbg{80.7}&\pbg{82.4}&\pbg{71.7}&\pbg{76.3}&\pbg{79.7}&\pbg{83}&\pbg{85}&\pbg{85.9}&\pbg{80}&\pbg{72.4}&\pbg{91.2}&\pbg{85}&\pbg{77.5}&\pbg{67.5}&\pbg{75.6}&\pbg{85}&\pbg{87.5}&\pbg{70}&\fbg{57.5}&\pbg{69.4}&\pbg{89.5}&\pbg{68}&27
\\
\hline
gpt-4o-2024-11-20&\pbg{83.8}&\fbg{63.7}&\pbg{75}&\pbg{77.5}&\fbg{59}&\fbg{55.6}&\pbg{74.6}&\pbg{82.7}&\pbg{78.4}&\pbg{84.1}&\pbg{75}&\pbg{78.8}&\pbg{87.7}&\pbg{76.6}&\pbg{86.2}&\pbg{91.3}&\pbg{80}&\pbg{71.4}&\pbg{91.2}&\pbg{80}&\pbg{72.5}&\pbg{68.8}&\pbg{75.6}&\pbg{87.5}&\pbg{87.5}&\pbg{67.5}&\fbg{55}&\pbg{66.7}&\pbg{81.6}&\pbg{68}&26
\\
\hline
gpt-4o-2024-08-06&\pbg{86.2}&\pbg{71.2}&\pbg{70}&\pbg{75}&\fbg{53.8}&\fbg{48.1}&\pbg{76.1}&\pbg{80}&\pbg{81.8}&\pbg{84.1}&\pbg{68.3}&\pbg{78.8}&\pbg{79.7}&\pbg{83}&\pbg{85}&\pbg{88}&\pbg{77.5}&\pbg{72.4}&\pbg{91.2}&\pbg{82.5}&\pbg{75}&\pbg{67.5}&\pbg{73.3}&\pbg{85}&\pbg{90}&\pbg{67.5}&\fbg{57.5}&\pbg{69.4}&\pbg{89.5}&\pbg{68}&27
\\
\hline
gpt-4o-2024-05-13&\pbg{91.2}&\pbg{66.2}&\pbg{75}&\pbg{76.2}&\pbg{66.7}&\fbg{53.8}&\pbg{78.3}&\pbg{82.7}&\pbg{79.5}&\pbg{85.3}&\pbg{78.3}&\pbg{78.8}&\pbg{89.9}&\pbg{78.7}&\pbg{82.5}&\pbg{93.5}&\pbg{80}&\pbg{69.4}&\pbg{88.8}&\pbg{90}&\pbg{77.5}&\pbg{68.8}&\pbg{73.3}&\pbg{85}&\pbg{87.5}&\pbg{68.8}&\fbg{52.5}&\pbg{66.7}&\pbg{89.5}&\pbg{68}&28
\\
\hline
chatgpt-4o-latest&\pbg{86.2}&\pbg{66.2}&\pbg{75}&\pbg{75}&\fbg{59.6}&\pbg{70}&\pbg{73.9}&\pbg{86.7}&\pbg{73.9}&\pbg{82.4}&\pbg{71.7}&\pbg{81.4}&\pbg{83.3}&\pbg{78.7}&\pbg{86.2}&\pbg{97.8}&\pbg{87.5}&\pbg{73.5}&\pbg{92.5}&\pbg{82.5}&\pbg{80}&\pbg{71.2}&\pbg{74.4}&\pbg{90}&\pbg{92.5}&\pbg{70}&\fbg{60}&\pbg{69.4}&\pbg{86.8}&\pbg{70.2}&28
\\
\hline
gpt-4o-mini-2024-07-18&\pbg{78.8}&\fbg{55}&\pbg{65}&\pbg{72.5}&\fbg{50.6}&\fbg{51.2}&\pbg{65.9}&\pbg{74}&\pbg{72.2}&\pbg{71.8}&\fbg{58.3}&\fbg{49.4}&\pbg{79.7}&\fbg{63.8}&\pbg{81.2}&\pbg{87}&\pbg{68.8}&\pbg{76.5}&\pbg{78.8}&\pbg{67.5}&\fbg{62.5}&\fbg{52.5}&\pbg{75.6}&\pbg{70}&\pbg{70}&\pbg{68.8}&\fbg{55}&\fbg{52.8}&\fbg{57.9}&\pbg{67}&19
\\
\hline
gpt-4-turbo&\pbg{83.8}&\pbg{72.5}&\pbg{70}&\pbg{73.8}&\pbg{65.4}&\fbg{41.2}&\pbg{75.4}&\pbg{78.7}&\pbg{69.3}&\pbg{84.1}&\pbg{75}&\fbg{62.2}&\pbg{80.4}&\pbg{72.3}&\pbg{81.2}&\pbg{83.7}&\pbg{82.5}&\pbg{73.5}&\pbg{86.2}&\pbg{80}&\pbg{77.5}&\pbg{65}&\pbg{72.1}&\pbg{77.5}&\pbg{85}&\pbg{67.5}&\fbg{50}&\pbg{72.2}&\pbg{78.9}&\pbg{76.7}&27
\\
\hline
gpt-4-turbo-2024-04-09&\pbg{81.2}&\pbg{67.5}&\pbg{67.5}&\pbg{78.8}&\pbg{65.4}&\fbg{48.8}&\pbg{75.4}&\pbg{78.7}&\pbg{69.3}&\pbg{77.6}&\pbg{75}&\fbg{62.2}&\pbg{80.4}&\pbg{74.5}&\pbg{83.8}&\pbg{81.5}&\pbg{82.5}&\pbg{73.5}&\pbg{91.2}&\pbg{82.5}&\pbg{80}&\pbg{65}&\pbg{72.1}&\pbg{77.5}&\pbg{85}&\pbg{70}&\fbg{52.5}&\pbg{69.4}&\pbg{78.9}&\pbg{68}&27
\\
\hline

\multicolumn{31}{l}{\textit{\textbf{Gemini Models}}}\\
gemini-3-flash-preview-result&\pbg{96.2}&\pbg{92.5}&\pbg{95}&\pbg{95}&\pbg{98.1}&\pbg{89.4}&\pbg{79.7}&\pbg{90.7}&\pbg{79.5}&\pbg{88.2}&\pbg{88.3}&\pbg{89.1}&\pbg{86.2}&\pbg{91.5}&\pbg{95}&\pbg{97.8}&\pbg{90}&\pbg{75.5}&\pbg{97.5}&\pbg{97.5}&\pbg{75}&\pbg{75}&\pbg{83.7}&\pbg{87.5}&\pbg{100}&\pbg{76.2}&\pbg{67.5}&\pbg{77.8}&\pbg{92.1}&\pbg{89.8}&30
\\
\hline
gemini-2.5-pro&\pbg{93.8}&\pbg{90}&\pbg{90}&\pbg{92.5}&\pbg{96.8}&\pbg{85.6}&\pbg{76.8}&\pbg{84}&\pbg{81.8}&\pbg{91.8}&\pbg{91.7}&\pbg{85.9}&\pbg{88.4}&\pbg{87.2}&\pbg{90}&\pbg{94.6}&\pbg{87.5}&\pbg{82.7}&\pbg{95}&\pbg{92.5}&\pbg{87.5}&\pbg{83.8}&\pbg{77.9}&\pbg{85}&\pbg{92.5}&\pbg{75}&\fbg{62.5}&\pbg{69.4}&\pbg{92.1}&\pbg{78.3}&29
\\
\hline
gemini-2.5-flash&\pbg{90}&\pbg{81.2}&\pbg{80}&\pbg{87.5}&\pbg{67.3}&\pbg{67.5}&\pbg{69.6}&\pbg{85.3}&\pbg{81.8}&\pbg{85.3}&\pbg{71.7}&\pbg{78.8}&\pbg{82.6}&\pbg{85.1}&\pbg{90}&\pbg{92.4}&\pbg{90}&\pbg{80.6}&\pbg{97.5}&\pbg{90}&\pbg{80}&\pbg{65}&\pbg{69.8}&\pbg{80}&\pbg{85}&\fbg{62.5}&\pbg{72.5}&\pbg{69.4}&\pbg{86.8}&\pbg{69.6}&29
\\
\hline
gemini-2.5-flash-lite&\pbg{83.8}&\fbg{63.7}&\pbg{67.5}&\pbg{73.8}&\fbg{49.4}&\fbg{45}&\pbg{78.3}&\pbg{81.3}&\pbg{77.3}&\pbg{79.4}&\pbg{81.7}&\fbg{59}&\pbg{79.7}&\pbg{78.7}&\pbg{87.5}&\pbg{97.8}&\pbg{80}&\pbg{76.5}&\pbg{82.5}&\pbg{85}&\pbg{70}&\pbg{71.2}&\pbg{75.6}&\pbg{77.5}&\pbg{77.5}&\fbg{62.5}&\fbg{40}&\fbg{58.3}&\pbg{78.9}&\fbg{59.8}&22
\\
\hline
gemini-2.5-flash-lite-preview-09-2025&\pbg{82.5}&\pbg{76.2}&\pbg{67.5}&\pbg{78.8}&\fbg{50}&\fbg{49.4}&\pbg{79}&\pbg{80}&\pbg{73.9}&\pbg{80.6}&\fbg{63.3}&\fbg{57.7}&\pbg{76.1}&\pbg{78.7}&\pbg{77.5}&\pbg{94.6}&\pbg{76.2}&\pbg{73.5}&\pbg{83.8}&\pbg{70}&\pbg{72.5}&\pbg{73.8}&\pbg{80.2}&\pbg{82.5}&\pbg{82.5}&\pbg{65}&\fbg{52.5}&\fbg{58.3}&\pbg{84.2}&\pbg{71.3}&24
\\
\hline
gemini-2.0-flash&\pbg{85}&\pbg{72.5}&\pbg{70}&\pbg{65}&\fbg{61.5}&\fbg{55.6}&\pbg{77.5}&\pbg{74.7}&\pbg{78.4}&\pbg{84.1}&\pbg{75}&\fbg{62.8}&\pbg{76.1}&\pbg{76.6}&\pbg{78.8}&\pbg{92.4}&\pbg{80}&\pbg{84.7}&\pbg{91.2}&\pbg{80}&\pbg{80}&\pbg{76.2}&\pbg{74.4}&\pbg{80}&\pbg{75}&\pbg{67.5}&\fbg{60}&\fbg{61.1}&\pbg{65.8}&\pbg{78.3}&25
\\
\hline
gemini-2.0-flash-lite&\pbg{77.5}&\pbg{67.5}&\pbg{75}&\pbg{71.2}&\fbg{53.2}&\fbg{41.9}&\pbg{70.3}&\pbg{76}&\pbg{68.2}&\pbg{80.6}&\pbg{68.3}&\fbg{63.5}&\pbg{76.8}&\pbg{72.3}&\pbg{82.5}&\pbg{93.5}&\pbg{85}&\pbg{72.4}&\pbg{85}&\pbg{75}&\pbg{77.5}&\pbg{75}&\pbg{80.2}&\pbg{75}&\pbg{77.5}&\fbg{62.5}&\fbg{52.5}&\fbg{63.9}&\pbg{71.1}&\pbg{81.1}&24
\\
\hline

\multicolumn{31}{l}{\textit{\textbf{Anthropic}}}\\
claude-3-haiku-20240307&\fbg{61.3}&\fbg{48.8}&\fbg{57.5}&\fbg{63.7}&\fbg{42.3}&\fbg{49.4}&\pbg{66.7}&\pbg{78.7}&\fbg{48.9}&\pbg{67.1}&\fbg{61.7}&\pbg{65.4}&\pbg{79}&\fbg{48.9}&\fbg{60}&\fbg{60.9}&\pbg{70}&\fbg{63.3}&\pbg{77.5}&\pbg{72.5}&\fbg{60}&\fbg{62.5}&\pbg{72.1}&\pbg{80}&\pbg{72.5}&\fbg{50}&\fbg{47.5}&\fbg{61.1}&\fbg{44.7}&\fbg{51.3}&11
\\
\hline
claude-haiku-4-5-20251001&\pbg{76.2}&\fbg{61.3}&\pbg{70}&\pbg{70}&\fbg{46.8}&\fbg{49.4}&\pbg{83.3}&\pbg{81.3}&\pbg{74.4}&\pbg{76.5}&\pbg{71.7}&\fbg{63.5}&\pbg{81.2}&\pbg{85.1}&\pbg{86.2}&\pbg{94.6}&\pbg{85}&\pbg{76.5}&\pbg{83.8}&\pbg{71.8}&\pbg{75}&\pbg{77.5}&\pbg{84.9}&\pbg{77.5}&\pbg{87.5}&\pbg{65}&\fbg{60}&\pbg{77.8}&\pbg{68.4}&\fbg{60.4}&24
\\
\hline
claude-opus-4-5-20251101&\pbg{88.8}&\pbg{77.5}&\pbg{82.5}&\pbg{86.2}&\fbg{64.1}&\pbg{71.9}&\pbg{79.7}&\pbg{87.3}&\pbg{87.5}&\pbg{86.5}&\pbg{81.7}&\pbg{77.6}&\pbg{84.8}&\pbg{87.2}&\pbg{91.2}&\pbg{95.7}&\pbg{85}&\pbg{81.6}&\pbg{100}&\pbg{92.5}&\pbg{85}&\pbg{83.8}&\pbg{89.5}&\pbg{80}&\pbg{97.5}&\pbg{70}&\fbg{60}&\pbg{69.4}&\pbg{92.1}&\pbg{83.3}&28
\\
\hline
claude-sonnet-4.5-result&\pbg{75}&\pbg{71.2}&\pbg{76.7}&\pbg{81.2}&\fbg{58.1}&\fbg{64.2}&\pbg{92.8}&\pbg{82.7}&\pbg{72.7}&\pbg{76.9}&\pbg{71.7}&\pbg{77.8}&\pbg{86.2}&\pbg{75.9}&\pbg{91.2}&\pbg{92}&\pbg{79.2}&\pbg{78.2}&\pbg{95.8}&\pbg{82.5}&\pbg{77.5}&\pbg{76.2}&\pbg{82.6}&\pbg{82.5}&\pbg{87.5}&\fbg{62.5}&\fbg{60}&\pbg{71.3}&\pbg{78.9}&\pbg{76.7}&26

\end{tabular}
\label{table:RQ1:result-1}
\end{table*}

\begin{table*}[t]

\caption{Results of Open-source Language models on different ISTQB Exams. On x-axis are the certification exam, on y-axis are the models; each cell shows the percentage scored by the model}
\label{tab:models-istqb-2}
\renewcommand{\arraystretch}{1.15}
\setlength{\tabcolsep}{1.95pt}
\scriptsize
\centering
\renewcommand{\arraystretch}{1.1}
\rowcolors{2}{white}{gray!10}
\begin{tabular}{p{2cm}cccccccccccccccccccccccccccccccc}
\toprule
\textbf{Model} &
\rotatebox{90}{CTFL - A} &
\rotatebox{90}{CTFL - B} &
\rotatebox{90}{CTFL - C} &
\rotatebox{90}{CTFL - D} &
\rotatebox{90}{CTAL-TAv4.1} &
\rotatebox{90}{CTAL-TAv3.1} &
\rotatebox{90}{CTAL-TAEv2.0} &
\rotatebox{90}{CTAL-TAEv1.3} &
\rotatebox{90}{CTAL-TMv3.0} &
\rotatebox{90}{CTAL-TMv1.4} &
\rotatebox{90}{CTAL-ATT} &
\rotatebox{90}{CTAL-TTA} &
\rotatebox{90}{CT-ATLaS} &
\rotatebox{90}{CT-AI} &
\rotatebox{90}{CTFL-AT} &

\rotatebox{90}{CT-GenAI} &
\rotatebox{90}{CT-MAT} &
\rotatebox{90}{CT-TAS} &
\rotatebox{90}{CT-MBT} &

\rotatebox{90}{CT-AcT} &
\rotatebox{90}{CT-PT} &

\rotatebox{90}{CT-SEC} &
\rotatebox{90}{CT-STE} &
\rotatebox{90}{CT-UT} &
\rotatebox{90}{CT-AuT} &
\rotatebox{90}{CT-GaMe} &

\rotatebox{90}{CT-GT} &
\rotatebox{90}{CTEL-ITP-ATP} &
\rotatebox{90}{CTEL-ITP-ITPI} &
\rotatebox{90}{CTEL-TM-OTM} & \rotatebox{90}{Total Passed} 

\\
\midrule

\\
\hline
\multicolumn{31}{l}{\textit{\textbf{Meta}}}\\
Llama 4 Maverick&\pbg{82.9}&\pbg{70.8}&\pbg{78.3}&\pbg{67.5}&\fbg{58.1}&\fbg{51.1}&\pbg{78.3}&\pbg{84}&\pbg{71.2}&\pbg{80}&\fbg{55.6}&\pbg{69.7}&\pbg{82.9}&\pbg{80.1}&\pbg{74.2}&\pbg{88.4}&\pbg{77.5}&\pbg{75.9}&\pbg{86.7}&\pbg{70.8}&\pbg{70}&\fbg{62.9}&\pbg{71.7}&\pbg{72.5}&\pbg{80.8}&\pbg{66.7}&\fbg{59.2}&\pbg{68.5}&\pbg{81.6}&\pbg{67.3}&25
\\
\hline
\multicolumn{31}{l}{\textit{\textbf{Mistral AI}}}\\
mistral-small3.2&\pbg{75}&\fbg{61.3}&\fbg{56.2}&\fbg{63.7}&\fbg{46.8}&\fbg{38.8}&\pbg{69.6}&\pbg{78.7}&\pbg{73.3}&\pbg{71.2}&\pbg{68.3}&\fbg{54.5}&\pbg{67.4}&\pbg{72.3}&\pbg{80}&\pbg{88}&\pbg{72.5}&\pbg{68.4}&\pbg{71.2}&\pbg{72.5}&\pbg{70}&\pbg{76.2}&\pbg{69.8}&\pbg{82.5}&\pbg{75}&\fbg{62.5}&\fbg{52.5}&\pbg{69.4}&\pbg{68.4}&\fbg{57.8}&21
\\
\hline

mistral-small3.1:latest&\pbg{73.8}&\pbg{71.2}&\fbg{61.3}&\fbg{62.5}&\fbg{48.1}&\fbg{51.9}&\pbg{70.3}&\pbg{73.3}&\pbg{73.9}&\pbg{75.3}&\pbg{78.3}&\fbg{63.5}&\pbg{65.9}&\fbg{63.8}&\pbg{81.2}&\pbg{92.4}&\pbg{75}&\pbg{72.4}&\pbg{73.8}&\pbg{75}&\pbg{77.5}&\pbg{80}&\pbg{80.2}&\pbg{72.5}&\pbg{77.5}&\pbg{67.5}&\fbg{47.5}&\pbg{75}&\fbg{55.3}&\fbg{61.5}&21
\\
\hline
ministral-3:3b&\pbg{67.5}&\fbg{41.2}&\fbg{43.8}&\fbg{48.8}&\fbg{46.2}&\fbg{41.2}&\fbg{59.4}&\pbg{66}&\pbg{67.6}&\fbg{54.1}&\fbg{60}&\fbg{49.4}&\fbg{54.3}&\fbg{55.3}&\pbg{67.5}&\pbg{90.2}&\fbg{58.8}&\fbg{53.1}&\pbg{67.5}&\pbg{67.5}&\fbg{62.5}&\fbg{51.2}&\pbg{72.1}&\pbg{67.5}&\fbg{60}&\fbg{53.8}&\fbg{50}&\fbg{63.9}&\pbg{65.8}&\fbg{43.7}&10
\\
\hline
ministral-3:8b&\pbg{71.2}&\fbg{53.8}&\fbg{62.5}&\pbg{65}&\fbg{47.4}&\fbg{53.8}&\fbg{58}&\pbg{74}&\fbg{64.2}&\pbg{69.4}&\fbg{58.3}&\fbg{59}&\pbg{70.3}&\fbg{63.8}&\pbg{77.5}&\pbg{80.4}&\pbg{72.5}&\fbg{60.2}&\pbg{65}&\pbg{70}&\pbg{70}&\fbg{60}&\pbg{67.4}&\pbg{65}&\pbg{80}&\pbg{65}&\fbg{52.5}&\pbg{72.2}&\pbg{65.8}&\fbg{50.7}&17
\\
\hline
ministral-3:14b&\pbg{67.5}&\fbg{61.3}&\fbg{57.5}&\pbg{65}&\fbg{43.6}&\fbg{45.6}&\fbg{64.5}&\pbg{74.7}&\pbg{69.9}&\pbg{72.9}&\pbg{68.3}&\pbg{73.1}&\pbg{73.2}&\pbg{74.5}&\pbg{85}&\pbg{85.9}&\pbg{80}&\pbg{69.4}&\pbg{68.8}&\pbg{70}&\pbg{80}&\pbg{65}&\pbg{75.6}&\pbg{65}&\pbg{77.5}&\fbg{56.2}&\fbg{55}&\fbg{55.6}&\fbg{60.5}&\pbg{73.5}&21
\\
\hline
\multicolumn{31}{l}{\textit{\textbf{Nvidia}}}\\

Nemotron-Nano&\pbg{73.1}&\pbg{75.6}&\pbg{69.2}&\pbg{82.1}&\pbg{78.3}&\pbg{76.1}&\pbg{76.1}&\pbg{82.7}&\fbg{60.5}&\pbg{78}&\pbg{68.3}&\fbg{64.4}&\pbg{78.6}&\pbg{78.3}&\pbg{82.9}&\pbg{91.3}&\pbg{72.5}&\pbg{65.6}&\pbg{85.9}&\pbg{71.1}&\pbg{69.2}&\pbg{71.8}&\pbg{72.1}&\pbg{80}&\pbg{76.9}&\pbg{68.4}&\fbg{52.5}&\pbg{72.2}&\pbg{73.7}&\pbg{79}&27
\\
\hline
\multicolumn{31}{l}{\textit{\textbf{Alibaba cloud}}}\\

Qwen2.5-VL-72B-Instruct&\pbg{78.8}&\pbg{71.2}&\pbg{70}&\pbg{75}&\fbg{57.1}&\pbg{66.2}&\pbg{66.7}&\pbg{76.7}&\pbg{68.2}&\pbg{81.2}&\pbg{81.7}&\fbg{64.1}&\pbg{86.2}&\pbg{78.7}&\pbg{80}&\pbg{93.5}&\pbg{80}&\pbg{77.6}&\pbg{90}&\pbg{75}&\pbg{82.5}&\pbg{71.2}&\pbg{80.2}&\pbg{77.5}&\pbg{77.5}&\fbg{63.7}&\fbg{55}&\pbg{69.4}&\pbg{76.3}&\pbg{68.7}&26
\\
\hline
qwen2.5vl:32b&\pbg{72.5}&\pbg{68.8}&\pbg{66.2}&\pbg{72.5}&\fbg{50}&\fbg{50.6}&\pbg{68.8}&\pbg{78.7}&\pbg{69.3}&\pbg{78.8}&\pbg{65}&\pbg{66.7}&\fbg{58}&\pbg{74.5}&\pbg{86.2}&\pbg{94.6}&\pbg{80}&\pbg{74.5}&\pbg{70}&\fbg{62.5}&\pbg{80}&\pbg{68.8}&\pbg{74.4}&\pbg{77.5}&\pbg{85}&\fbg{62.5}&\pbg{72.5}&\pbg{66.7}&\fbg{60.5}&\pbg{79.6}&24
\\
\hline
qwen2.5vl:3b&\fbg{52.5}&\fbg{47.5}&\fbg{53.8}&\fbg{45}&\fbg{41.7}&\fbg{33.8}&\fbg{54.3}&\fbg{64}&\pbg{71.6}&\fbg{62.4}&\fbg{43.3}&\fbg{42.3}&\fbg{52.2}&\fbg{61.7}&\pbg{76.2}&\pbg{70.7}&\fbg{58.8}&\pbg{67.3}&\fbg{63.7}&\fbg{55}&\pbg{67.5}&\fbg{51.2}&\pbg{65.1}&\fbg{52.5}&\pbg{70}&\fbg{51.2}&\fbg{47.5}&\fbg{50}&\fbg{60.5}&\fbg{61.1}&7
\\
\hline
qwen2.5vl:7b&\pbg{67.5}&\fbg{60}&\fbg{55}&\fbg{53.8}&\fbg{35.3}&\fbg{52.5}&\pbg{69.6}&\pbg{75.3}&\fbg{63.1}&\fbg{59.4}&\pbg{68.3}&\fbg{48.7}&\fbg{58.7}&\fbg{57.4}&\pbg{67.5}&\pbg{75}&\pbg{72.5}&\pbg{70.4}&\pbg{66.2}&\pbg{67.5}&\pbg{67.5}&\fbg{58.8}&\fbg{62.8}&\fbg{57.5}&\pbg{77.5}&\fbg{62.5}&\fbg{55}&\pbg{72.2}&\pbg{65.8}&\fbg{50.2}&14
\\
\hline
\multicolumn{31}{l}{\textit{\textbf{Google}}}\\

gemma-3-27b-it&\pbg{73.8}&\fbg{63.7}&\pbg{67.5}&\pbg{67.5}&\fbg{46.8}&\fbg{49.4}&\pbg{73.9}&\pbg{76.7}&\pbg{79.5}&\pbg{79.4}&\fbg{55}&\pbg{66.7}&\pbg{82.6}&\pbg{74.5}&\pbg{85}&\pbg{88}&\pbg{77.5}&\fbg{55.1}&\pbg{78.8}&\pbg{75}&\pbg{75}&\pbg{70}&\pbg{76.7}&\pbg{82.5}&\pbg{85}&\pbg{66.2}&\fbg{60}&\fbg{61.1}&\pbg{76.3}&\pbg{72.4}&23
\\
\hline
gemma3:latest&\fbg{56.2}&\fbg{23.8}&\fbg{40}&\fbg{43.8}&\fbg{32.1}&\fbg{42.5}&\fbg{52.2}&\pbg{66}&\fbg{64.2}&\fbg{58.2}&\fbg{41.7}&\fbg{39.1}&\fbg{53.6}&\fbg{59.6}&\pbg{72.5}&\fbg{64.1}&\pbg{70}&\fbg{62.2}&\fbg{61.3}&\fbg{60}&\fbg{57.5}&\fbg{47.5}&\pbg{65.1}&\fbg{40}&\pbg{65}&\fbg{55}&\fbg{47.5}&\fbg{55.6}&\fbg{47.4}&\fbg{47.4}&5
\\
\hline
gemma3:12b&\pbg{75}&\pbg{66.2}&\fbg{61.3}&\pbg{71.2}&\fbg{36.5}&\fbg{32.5}&\pbg{68.1}&\pbg{74}&\pbg{69.3}&\pbg{69.4}&\fbg{51.7}&\pbg{65.4}&\fbg{64.5}&\pbg{66}&\pbg{80}&\pbg{81.5}&\pbg{77.5}&\fbg{62.2}&\pbg{76.2}&\fbg{60}&\pbg{70}&\pbg{75}&\pbg{73.3}&\pbg{70}&\pbg{80}&\pbg{66.2}&\fbg{55}&\pbg{66.7}&\pbg{65.8}&\fbg{53.5}&21
\\
\hline

\multicolumn{31}{l}{\textit{\textbf{IBM}}}\\
Granite3.2-vision:2b&\fbg{57.5}&\fbg{33.8}&\fbg{35}&\fbg{35}&\fbg{26.3}&\fbg{27.5}&\fbg{42.8}&\fbg{53.3}&\fbg{53.4}&\fbg{33.5}&\fbg{55}&\fbg{32.1}&\fbg{32.6}&\fbg{31.9}&\fbg{50}&\fbg{47.8}&\fbg{42.5}&\fbg{49}&\fbg{41.2}&\fbg{42.5}&\fbg{50}&\fbg{43.8}&\fbg{46.5}&\fbg{37.5}&\fbg{50}&\fbg{33.8}&\fbg{47.5}&\fbg{50}&\fbg{50}&\fbg{45.7}&0
\\
\hline
Granite3.3 vision:2b&\fbg{57.5}&\fbg{37.5}&\fbg{37.5}&\fbg{33.8}&\fbg{28.8}&\fbg{12.5}&\fbg{39.1}&\pbg{67.3}&\fbg{54.5}&\fbg{42.4}&\fbg{43.3}&\fbg{34.6}&\fbg{52.9}&\fbg{38.3}&\fbg{47.5}&\fbg{54.3}&\fbg{50}&\fbg{42.9}&\fbg{50}&\fbg{45}&\fbg{55}&\fbg{46.2}&\fbg{38.4}&\fbg{52.5}&\fbg{52.5}&\fbg{37.5}&\fbg{50}&\fbg{58.3}&\fbg{52.6}&\fbg{10.9}&1
\\
\hline
\multicolumn{31}{l}{\textit{\textbf{Misc}}}\\

llava:34b&\pbg{70}&\fbg{50}&\fbg{55}&\fbg{53.8}&\fbg{44.2}&\fbg{48.8}&\pbg{65.2}&\pbg{70}&\pbg{72.2}&\fbg{64.7}&\fbg{46.7}&\fbg{54.5}&\pbg{65.2}&\fbg{63.8}&\pbg{73.8}&\fbg{63}&\pbg{77.5}&\pbg{65.3}&\pbg{67.5}&\fbg{60}&\pbg{65}&\fbg{53.8}&\pbg{75.6}&\pbg{65}&\fbg{62.5}&\fbg{48.8}&\fbg{52.5}&\pbg{66.7}&\fbg{47.4}&\fbg{54.3}&13
\\
\hline
llava:7b&\fbg{48.8}&\fbg{43.8}&\fbg{32.5}&\fbg{50}&\fbg{35.3}&\fbg{36.9}&\fbg{55.1}&\fbg{60.7}&\fbg{54.5}&\fbg{51.8}&\fbg{46.7}&\fbg{60.9}&\fbg{47.8}&\fbg{46.8}&\fbg{53.8}&\pbg{65.2}&\fbg{51.2}&\fbg{46.9}&\fbg{55}&\fbg{45}&\fbg{62.5}&\fbg{33.8}&\fbg{39.5}&\fbg{45}&\fbg{45}&\fbg{40}&\fbg{42.5}&\fbg{61.1}&\fbg{57.9}&\fbg{33.7}&1
\\
\hline
bakllava:latest&\fbg{45}&\fbg{30}&\fbg{41.2}&\fbg{37.5}&\fbg{16}&\fbg{23.1}&\fbg{26.8}&\fbg{53.3}&\fbg{38.1}&\fbg{35.3}&\fbg{33.3}&\fbg{32.1}&\fbg{27.5}&\fbg{34}&\fbg{36.2}&\fbg{54.3}&\fbg{35}&\fbg{33.7}&\fbg{36.2}&\fbg{47.5}&\fbg{60}&\fbg{40}&\fbg{37.2}&\fbg{52.5}&\fbg{40}&\fbg{28.7}&\fbg{27.5}&\fbg{41.7}&\fbg{39.5}&\fbg{34.6}&0
\\
\hline
llava-phi3:latest&\fbg{57.5}&\fbg{42.5}&\fbg{46.2}&\fbg{50}&\fbg{38.5}&\fbg{42.5}&\fbg{45.7}&\fbg{62}&\fbg{59.7}&\fbg{57.6}&\fbg{55}&\fbg{44.9}&\fbg{52.2}&\fbg{57.4}&\fbg{63.7}&\fbg{63}&\pbg{70}&\fbg{48}&\fbg{58.8}&\fbg{47.5}&\fbg{60}&\fbg{40}&\pbg{68.6}&\fbg{50}&\pbg{65}&\fbg{51.2}&\fbg{35}&\fbg{52.8}&\fbg{52.6}&\fbg{53.9}&3
\\
\hline
minicpm-v:latest&\fbg{58.8}&\fbg{42.5}&\fbg{43.8}&\fbg{42.5}&\fbg{29.5}&\fbg{42.5}&\fbg{56.5}&\fbg{58.7}&\fbg{56.2}&\fbg{62.4}&\fbg{36.7}&\fbg{44.9}&\fbg{60.9}&\fbg{44.7}&\fbg{61.3}&\fbg{58.7}&\fbg{63.7}&\fbg{50}&\fbg{56.2}&\fbg{62.5}&\fbg{55}&\fbg{46.2}&\fbg{62.8}&\fbg{52.5}&\fbg{55}&\fbg{48.8}&\fbg{40}&\pbg{69.4}&\fbg{39.5}&\fbg{56.7}&1
\\
\hline
llava-llama3:latest&\fbg{56.2}&\fbg{33.8}&\fbg{51.2}&\fbg{42.5}&\fbg{28.8}&\fbg{44.4}&\fbg{41.3}&\fbg{58}&\fbg{58.5}&\fbg{41.8}&\fbg{60}&\fbg{46.2}&\fbg{40.6}&\fbg{40.4}&\fbg{56.2}&\fbg{44.6}&\fbg{43.8}&\fbg{48}&\fbg{51.2}&\fbg{37.5}&\fbg{57.5}&\fbg{50}&\fbg{43}&\fbg{50}&\fbg{47.5}&\fbg{38.8}&\fbg{40}&\fbg{47.2}&\fbg{42.1}&\fbg{37}&0
\\
\hline

\bottomrule

\end{tabular}
\label{table:RQ1:result-2}
\end{table*}
We use all the publicly available distinct sample exams for software testing certifications from ISTQB. 
Table~\ref{table:ISTQBExam} shows the details of all of the available certifications by ISTQB.
We excluded  CTEL-TM-SM,  CTEL-TM-MTT from our dataset because their sample exam is same as CTEL-TM-OTM. 
As described in section~\ref{sec:background:ISTQB_certification_exam}, ISTQB exams have different levels of questions in the exams. 
In total, each language model has to solve 1171 questions: 31 questions with images, 114 questions of level K1, 667 questions of level K2, 245 questions of level K3, and 135 questions of level K4. Table~\ref{table:klevel-counts} shows the distribution of different levels of questions for each test.
For the complete experiments, we executed 71,290 prompts (1171 questions x 60 language models + 515 transformed questions x 2 language models) on language models for the certification exam.
\begin{table*}[htbp]
\centering
\footnotesize
\caption{Summary of ISTQB Certification Exams (Foundation, Advanced, Specialist and Expert)}
\label{table:istqb-summary}
\renewcommand{\arraystretch}{1.1}
\rowcolors{2}{white}{gray!10}
\begin{tabular}{p{7cm}p{2.5cm}cp{1.8cm}p{2cm}}
\toprule
\textbf{Exam Name (Acronym)} & 
\textbf{Level/Type} & 
\textbf{Questions} & 
\textbf{Total Points} & 
\textbf{Passing (65\%)} \\
\midrule
Certified Tester Foundation Level v4.0 (CTFL) & Core Foundation & 40 & 40 & 26 points \\

\midrule
Certified Tester Advanced Level Test Analyst (CTAL-TA) v4.0 & Core Advanced & 45 & 78 & 51 points \\
Certified Tester Advanced Level Test Automation Engineering (CTAL-TAE) v2.0  & Core Advanced & 40 & 66 & 43 points \\

Certified Tester Advanced Level Test Management (CTAL-TM) v3.0 & Core Advanced & 50 & 88 & 58 points \\
Certified Tester Advanced Level Technical Test Analyst (CTAL-TTA) & Core Advanced & 45 & 78 & 51 points \\

Certified Tester Advanced Level Test Analyst (CTAL-TA) - old  & Core Advanced & 40 & 80 & 52 points \\

Certified Tester Advanced Level Test Manager (CTAL-TM) 2012 & Core Advanced & 65 & 115 & 75 points \\

\midrule
\multicolumn{5}{l}{\textit{\textbf{Technologies and Approaches}}} \\
Certified Tester Agile Technical Tester (CTAL-ATT) & Specialist & 40 & 64 & 42 points \\

Certified Tester Agile Test Leadership at Scale
(CT-ATLaS) & Specialist & 40 & 71 & 47 points \\

Certified Tester AI Testing (CT-AI) & Specialist & 45 & 47 & 31 points \\

Certified Tester Foundation Level Agile Tester (CTFL-AT) & Specialist & 40 & 40 & 26 points \\

Certified Tester Testing with Generative AI (CT-GenAI)  & Specialist & 40 & 46 & 30 points \\

Certified Tester Mobile Application Testing (CT-MAT) & Specialist & 40 & 40 & 26 points \\

Certified Tester Test Automation Strategy (CT-TAS) & Specialist & 40 & 49 & 32 points \\
Certified Tester Model-Based Tester (CT-MBT) & Specialist & 40 & 40 & 26 points \\

Certified Tester Test Automation Engineer (CT-TAE) 2016 & Specialist & 40 & 75 & 49 points \\
\multicolumn{5}{l}
{\textit{\textbf{Quality Characteristics \& Test Levels}}} \\

Certified Tester Acceptance Testing (CT-AcT) & Specialist & 40 & 40 & 26 points \\

Certified Tester Performance Testing (CT-PT) & Specialist & 40 & 40 & 26 points \\

Certified Tester Security Test Engineer (CT-STE) & Specialist & 40 & 43 & 28 points \\

Certified Tester Security Tester (CT-SEC) & Specialist & 45 & 80 & 52 points \\
Certified Tester Usability Testing (CT-UT) & Specialist & 40 & 40 & 26 points \\
\multicolumn{5}{l}{\textit{\textbf{Testing in Particular Domains}}} \\
Certified Tester Automotive Software Tester (CT-AuT) & Specialist & 40 & 40 & 26 points \\

Certified Tester Game Testing (CT-GaMe)
 & Specialist & 40 & 40 & 26 points \\
 
Certified Tester Gambling Industry Tester (CT-GT)  & Specialist & 40 & 40 & 26 points \\

\midrule
\multicolumn{5}{l}{\textit{\textbf{Improving the Test Process}}} \\

Certified Tester Expert Level Assessing Test Processes (CTEL-ITP-ATP) & Expert & 25 (only MCQs) & 36 & 23 points \\
Certified Tester Expert Level Implementing Test Process Improvement (CTEL-ITP-ITPI) & Expert & 25 (only MCQs) & 40 & 26 points \\
\multicolumn{5}{l}{\textit{\textbf{Test Management}}} \\

Certified Tester Expert Level Test Management Operational Test Management (CTEL-TM-OTM) & Expert & 20 (only MCQs) & 46 & 29 points \\

Certified Tester Expert Level Test Management Strategic Test Management (CTEL-TM-SM)& Expert & 20 (only MCQs) & 46 & 29 points \\

Certified Tester Expert Level Test Management Managing the Test Team (CTEL-TM-MTT) & Expert & 20 (only MCQs) & 46 & 29 points \\

\bottomrule
\end{tabular}
\label{table:ISTQBExam}

\end{table*}
\subsubsection{Experimental Setup and Hyperparameters}
\hfill\\
Commercial LLMs are accessed externally via an API key. The only exception is Claude Sonnet 4.5. Claude Sonnet and LLama Maverick are accessed via an AI assistant deployed on Azure AI Foundry.
We evaluated smaller open-source models (<12B parameters) on Ubuntu 20.04 running on i-7, with RTX 4070 8GB and 16GB memory.
For Qwen2.5 with 72 billion parameters and Nvidia Nemotron nano 2, we accessed the models through the Nebius Token Factory API~\cite{nebius_token_factory}.
The remaining open-source models were executed on Windows 11 running on an Intel 12th Gen i7 CPU with 32 GB memory. 
For the models that allow adjusting the parameters, we set the temperature to 0 and kept the top\_p to 1 (default value) to have the deterministic output.
For the models that do not allow changing the parameters, such as GPT-5, GPT-5 mini, GPT-5 nano etc. We execute them three times and report the average to account for the randomness in the output.

\subsection{RQ1: \textit{How do different language models fare in terms of passing the certification exams? }} 
\subsubsection{Results} \label{RQ1:results}
Table~\ref{tab:models-istqb-1} and ~\ref{tab:models-istqb-2} show the overall results of commercial and open-source language models, respectively, we evaluated in this study. 
On the x-axis are all the tests used in the evaluation. 
Note that, some certifications include multiple exams for one of two reasons: (1) multiple sample exams exist for the same certification such as CTFL, which has four sample exams available on ISTQB website, or (2) ISTQB provides sample different exams corresponding to different syllabi for the same certification, with one being outdated and the other current such as CTAL-TMv1.4 and CTAL-TMv3.0.
On the y-axis are all the models used in the evaluation. Note that some models have a timestamp concatenated in the name, this timestamp shows the date on which that model was released. 
Each cell shows the percentage for the model on the exam. The green background shows that the model passed (i.e., achieved more than 65\%) the exam and the red shows it failed. 
We calculate the marks by comparing the number of correct options given by the LLM. For example, if a question requires two answers and a language model provides one answer correctly, it receives half of the total marks for that question.

We first report the comparison of all language models collectively, followed by a comparison of models with respect to their sizes, and finally, we perform a focused comparison of open-source models. 
GPT-5-2025-08-7 and Gemini-3-flash-preview outperform other language models by scoring 65\% in all of the certification exams.
Overall, LLMs from commercial vendors outperform the models open-source community. None of the open-source models were able to pass all of the certification exams.
Nemotron-Nano, top performing open-source language model, is able to pass 27 out of 30 exams.

We also compared mini-sized and nano-sized models. The mini-sized category includes all mini models released by OpenAI. Furthermore, this category also incorporates the lite version of models from Gemini. 
In addition, it also includes medium-sized open-source models. Specifically, this category includes the models with the number of parameters ranging from 3 billion to 15 billion. 
The nano-sized category includes all the nano models from OpenAI. Furthermore, it also includes open-source models with parameters less than or equal to three billion.

In the mini category, GPT-5-mini, and GPT-5-mini-2025-08-07 perform better than the other models of the same size by passing on 29 out of 30 exams.  
On the second rank in the mini category is O4-mini which passed 29 exams, and on the third is o4-mini-2025-04-16 which passed 28 exams.
In the nano-category, GPT-5-nano and GPT-5-nano-2025-08-07 both passed 25 out of 30 exams. On the second rank is GPT-4.1-nano, which passed only 18 exams, and on the third is GPT-4.1-nano-2025-04-14, which is able to pass only 16 exams.
It is evident from the results that in the case of nano-sized models, only GPT-5 variants can be used.
In open-source models, Nemotron-Nano outperform other models. Nemotran-Nano is able to pass 27 exams out of 30.
In the second rank is Qwen2.5vl with 72 billion parameters, which is able to pass 26 exams, and in the third rank is LLama-4-Maverick, which is able to pass only 25 exams. 

To conclude, Gemini-3-flash-preview and GPT-5-2025-08-07 performed better than the rest of the language models in passing the software testing certification exams. 
However, selecting the most suitable model for performing software testing tasks should also consider  other factors such as computational resources, budget constraints, and deployment needs.

While RQ1 assesses whether language models can pass the overall certification exams based on the final score, it does not reveal how the model perform across different learning objective levels. We therefore, investigate this performance in RQ2. 


\subsection{RQ2: \textit{How do different language models fare in terms of attempting questions with different learning objectives levels?}}

\subsubsection{Results}
Table~\ref{table:RQ2} shows the number of correct answers for each question level by each language model.
On the x-axis are the question levels (K1 to K4) and the total column.
On the y-axis are all the models used in the evaluation.
One cell shows the number of correct answers given by the language model for the particular question level. 
For example, 101 (2nd row, second column) shows that GPT-5.2 answered 101 questions correctly on level K1. In  Table~\ref{table:RQ2}, there are some values in decimal points, such as 104.3 for GPT-5-mini for K-1; these are the models that do not allow setting the temperature value. Such models are executed 3 times to account for randomness and the average is reported in the table. 

Similar to RQ1, first, we report the comparison of all the language models collectively, followed by the comparison of models according to their size, and then finally, the comparison between open source models.
It is evident from the results that none of the models was able to answer all questions correctly for any knowledge level.

For questions at the K1 level, which test the ability to recall or recognize a term, Gemini-3-flash-preview performs the best. Gemini-3-flash-preview is able to correctly answer 112 questions out of 114 ($\sim 98$\%). At the second rank, GPT-5-2025-08-07 is able to correctly answer 110.3 ($\sim 96$\%), and on the third, GPT-5 with 109 out of 114 ($\sim 95$\%).

For questions at the K2 level, which test the ability to understand or interpret a concept or topic. 
Gemini-3-flash-preview outperforms other language models. Gemini-3-flash-preview is able to correctly answer 592 questions out of 667 ($\sim 89$\%). At the second rank, Gemini-2.5-pro answers 587 questions correctly ($\sim 88$\%).  
At the third rank, GPT-5-2025-08-7 answers 572 questions correctly ($\sim 85$\%).

For questions at the K3 level, which test the ability to apply learned concepts. Gemini-3-flash-preview outperforms other language models. Gemini-3-flash-preview is able to correctly answer 210 questions out of 245 ($\sim 85$\%). At the second rank, GPT-5 answers 206 questions correctly ($\sim 84$\%).  At the third rank, Gemini-2.5-pro answers 201 answers correctly ($\sim 82$\%).

For questions at the K4 level, which test the ability to break down information for better understanding. 
Gemini-3-flash-preview and Gemini-2.5-pro
outperform other language models. Both are able to correctly answer 110 out of 135 ($\sim 82$\%). 
At the second rank, Claude-Opus-4-5 answers 106 questions correctly   ($\sim 78$\%). 
At the third rank, GPT-5, and O4-mini-2025-04-16 answer 104.3 questions correctly ($\sim 77$\%).  

For comparisons based on model size, we adopt the mini and nano categories as defined in Section~\ref{RQ1:results}.
In the mini category, GPT-5-mini-2025-08-07  achieves the highest performance, correctly answering 104.3 out of 114 K1-level questions ($\sim$91\%).
For K2-level questions, both GPT-5-mini and GPT-5-mini-2025-08-07 achieve 552.3 out of 667 correct answers ($\sim$82\%). 
For K3-level, GPT-5-mini scores 194 out of 245 ($\sim$79\%) and 96.7 out of 135 ($\sim$71\%) for K4.
Overall, GPT-5-mini and  GPT-5-mini-2025-08-07 outperform other models within the mini category.

In the nano category, GPT-5-nano variants consistently outperform other models. GPT-5-nano-2025-08-07 correctly answers 94.7 out of 114 K1-level questions ($\sim$83\%), and for K2 level questions, it achieves 499.3 out of 667 correct answers ($\sim$74\%). For K3, GPT-5-nano-2025-08-07 obtains 159.7 out of 245 ($\sim$65\%), and for K4, GPT-5-nano achieves 75.7 out of 135 ($\sim$56\%). These results demonstrate that the nano variants of GPT-5 maintain a consistent performance advantage over other language models across all knowledge levels in the nano category.

Among open-source models, Qwen2.5vl:32b parameters outperform others with 97 correct answers out of 114 for K1. For questions with K2 level, Qwen2.5vl:72b answers 535 questions correctly out of 667. Llama 4 Maverick answers 150.3 questions correctly out of 245 for K3 and Qwen2.5vl:72b answers 93 questions correctly out of 135. 
It is clear that Qwen2.5vl variants should be the choice of model among open source models when solving questions with levels K1, K2, and K4, while Llama 4 Maverick should be selected for questions with level K3. 
Clearly, commercial models perform better than the open-source models. As expected, the performance of the language model drops (i.e., the ability to correctly answer questions) as the question level becomes more difficult. 

To conclude, similar to results for RQ1, Gemini-3-flash-preview outperform other models at individual question levels by answering 87\% questions correctly.

For practitioners aiming to deploy LLMs for software testing tasks, these results provide practical recommendations. 
If the use of commercial platforms is permitted, depending upon the industry guidelines, Gemini-3-flash-preview is the most suitable choice. However, if the organizational policies or budget constraints prevent the use of commercial LLMs, Qwen2.5vl:32b is recommended for tasks at cognitive levels K1, K2, and K4, while Llama 4 Maverick is better suited for tasks at level K3.

\subsection{RQ3: \textit{How do best performing language models fare when ISTQB exam questions undergo metamorphic transformations, reflecting robust understanding rather than memorization?}}
Several studies have applied metamorphic transformation to text as it solves the oracle generation problem in evaluating language models~\cite{cho2025metamorphic} . 
To answer RQ3, we collect, for each model, one question of each different learning objective level (K1 to K4) from every sample exam that the model has answered correctly. 
We apply semantic-preserving metamorphic transformation on these questions, ensuring that the underlying question does not change. 
All transformed questions are manually verified to be semantic-preserving by a professional holding a core advanced ISTQB certification.

We chose the language models that pass all of the certification exams from RQ1. The rationale behind this is that these top-performing language models consistently achieve passing scores across all four categories (i.e, core foundation, core advanced, specialist and expert)  of ISTQB certifications. Focusing on these models allows us to investigate whether their success is due to memorization of specific questions or they truly understand the software testing principles. As only GPT-5-2025-08-07 and Gemini-3-flash-preview pass all the exams in RQ1, they are selected for RQ3.
 
We implement six metamorphic relations (MR) derived from the literature: (MR-1) synonym substitution~\cite{hyun2024metal} to verify whether language model understand the concept or merely memorized the question, (MR-2) adding a single neutral sentence~\cite{srinivasan2018quality} to test whether model can focus on the key concept of question, (MR-3) adding multiple neutral sentences~\cite{wang2023mttm} to further increase the question complexity to evaluate the model's capability to filter out noise, (MR-4) swapping order of possible options~\cite{li2024empirical} to test whether model's answer depends on the position of option,
(MR-5) adding irrelevant options~\cite{li2024empirical}, to test whether model can still select the correct option even with the presence of more distractors, and (MR-6) replacing all options with irrelevant one and including a "none of above" option to test whether model can recognize that none of the options provided are correct.
An example of MR-1 is, consider the question from CT-MAT: "Which of the following statements is true for feature phones?". This is transformed into "Which of the given claims is true for feature phones?". While the wording is altered, the underlying meaning of the question remains the same. 

With two models, six transformations, and 87 questions each. We generated 515 transformed questions for each language model.
Each transformed question is queried to the language model using the same pattern as in RQ1 and their response is recorded and compared with their original answer.
\subsubsection{Results}
The evaluation of two top performing models (i.e.,  GPT-5-2025-08-07 and Gemini-3-flash-preview) shows a drop in performance when queried with transformed questions.  GPT-5-2025-08-07 answered 482 out of 515 correctly (93.6\%) while Gemini-3-flash-preview answered 471 questions correctly (91.5\%).
MR-6, which replaces all the options with irrelevant options and includes a "none of above" option, triggered the largest number of incorrect answers for both models (12 (36\%) for GPT  and 8 (18\%) for Gemini).
Other transformations, such as swapping options and adding irrelevant options (MR-4 and MR-5), caused fewer incorrect answers for both models (4 and 5 for GPT, and 5 and 7 for Gemini, respectively).
In the case of MR-1 (i.e., synonym substitution), GPT answered only 3 questions incorrectly, while Gemini struggled with MR-1 as it answered 8 questions incorrectly. This shows that GPT is more resilient to minor changes in the question as compared to Gemini. 
Similarly, for MR-2 and MR-3 (i.e, adding one or multiple neutral sentences), GPT incorrectly answered only 4 and 5 questions, respectively, while Gemini answered  7 and 8 questions incorrectly. This again shows that Gemini is less robust to minor changes in the question as compared to GPT.

The result indicate that although top-performing models answered more than 90\% of the metamorphically transformed questions correctly which shows that these models possess strong understanding of software testing principle. However, minor changes to the questions causes them to select the wrong option, showing that their performance is not completely robust. 
These findings suggest that that even high-performing models can be influenced by small changes in the questions, highlighting the need for further research on improving the robustness of the language models.

\subsection{Analysis and Recommendations for Incorrectly Answered Exam Questions}

We performed a question-level analysis of all the questions that were answered incorrectly by all the 60 models in our study. 

In total, we identified 15 questions that were answered incorrectly by all of the language models. We found commonalities in the questions.
First, the majority of the identified questions  requires selecting the most appropriate or best possible answer to question rather than objectively correct one. For example, "Which TWO of the following statements does BEST describe the impact of a software development model on security testing?" Language models are unable to prioritize the options without subtle cues or context.

Second, the presence of very similar distractors (i.e., options apart from the correct answer) is another major cause of answering the question incorrectly. For example, consider the question in the CT-GaMe exam ``Which of the following is an example of the typical input devices (guaranteed to comply with gaming devices)?'', the options are  ``a) Microphone, b) Steering wheel, c) Touch screen, and d) Joystick''. 
According to the answer sheet provided by ISTQB, the correct answer for this question is ``c'' because ``Touch screens are used in phones, PDAs, portable consoles, and modern slot machines and considered as typical input devices''. 
While this is correct within the broader context of gaming technology, the presence of such plausible distractors can make the question ambiguous. 
The distractors in this question can also be recognized as input devices commonly associated with the gaming industry. This type of ambiguity can confuse both machines and humans.

Third, several incorrectly answered questions are from a specific domain, such as casino systems or gaming devices. Such questions requires the domain specific knowledge which can be underrepresented in general-purpose training data of language models.

Lastly, questions that include multiple reasoning steps are also part of the subset of questions answered incorrectly. For example, consider the question "Consider the following technical environments: 1. Single computer 2. Multi-tier system 3. Distributed 4. Virtualized 5. Dynamic/Cloud-based 6. Client/Server and Browser-based 7. Mobile 8. Embedded 9. Mainframe Which of these is most likely to have a performance risk due to connectivity issues?" It requires the language model to first understand the different technical environments, then understand its role in performance and how it will be affected due to connectivity issues. This multi-stage reasoning increases the complexity of the question, and as a result language models struggle to answer such questions correctly.

Based on these findings, we recommend that questions should be designed to minimize ambiguity. 
In particular, questions that require selecting the best possible or most appropriate options should contain enough information to help distinguish between multiple options. 
Furthermore, when formulating distractors, we recommend that the distractors should be conceptually different than the correct one rather than providing equally plausible options in the same context.
In addition, the domain-specific questions should contain enough information that the examinee relies on the information available in the question rather than familiarity with the domain.
Finally, multi-step reasoning questions should include all the necessary information for making a decision without requiring the unstated assumptions.
These refinements can improve the clarity and fairness in the exams for both human candidates and language models.

\subsection{Replication Package}
The replication package of our experiments is available on Figshare~\cite{supp}.
It includes: (i) scripts to parse the PDF files for extracting questions, answers, question level and points for each question, (ii) scripts to query the open-source language model, (iii) scripts to calculate results, (iv) scripts to perform the metamorphic transformations (v) scripts to extract the question, answered correctly, of each learning level and (vi) readme file to setup the experiment.



\begin{table}
\centering
\caption{Number of Correct Answers for Each Question Level by Each Language Model}
\label{table:RQ2}
\footnotesize
\begin{tabular}{{p{3.3cm}lllll}}

\textbf{Model Name} & \textbf{K1} & \textbf{K2} & \textbf{K3} & \textbf{K4} & \textbf{Total} \\
\hline
gpt-5.2&101 & 553 & 172 & 101&927 
 \\ 
 \hline 
gpt-5.1&106 & 550 & 165 & 99&920 
 \\ 
 \hline 

 gpt-5&109.0&566.0&206.0&104.3&985.3 
 
\\ 
 \hline 
gpt-5-mini&103.3&552.3&194.0&96.7&946.3 
 \\ 
 \hline 
gpt-5-nano&94.3&499.7&158.0&75.7&827.7 
 \\ 
 \hline 
gpt-5-2025-08-07&110.3&572.0&198.3&102.3&983.0 
 \\ 
 \hline 
gpt-5-mini-2025-08-07&104.3&552.3&187.7&95.7&940.0 
 \\ 
 \hline 
gpt-5-nano-2025-08-07&94.7&499.3&159.7&73.7&827.3 
 \\ 
 \hline 
gpt-5-chat-latest&104 & 555 & 175 & 97&931 
 \\ 
 \hline 
gpt-4.1&104.0&557.0&164.0&96.0&921.0 
 \\ 
 \hline 
gpt-4.1-mini&97 & 526 & 155 & 95&873 
 \\ 
 \hline 
gpt-4.1-nano&84 & 451 & 131 & 74&740 
 \\ 
 \hline 
gpt-4.1-2025-04-14&104 & 557 & 164 & 98&923 
 \\ 
 \hline 
gpt-4.1-mini-2025-04-14&98 & 524 & 153 & 95&870 
 \\ 
 \hline 
gpt-4.1-nano-2025-04-14&84 & 446 & 132 & 72&734 
 \\ 
 \hline 
o4-mini&104.0&542.3&188.3&102.3&937.0 
 \\ 
 \hline 
o4-mini-2025-04-16&103.3&538.7&189.0&104.3&935.3 
 \\ 
 \hline 
o3&106.7&553.7&185.3&90.0&935.7 
 \\ 
 \hline 
o3-2025-04-16&104.3&551.3&181.3&91.3&928.3 
 \\ 
 \hline 
gpt-4o&100.0&535.0&147.0&95.0&877.0 
 \\ 
 \hline 
gpt-4o-2024-11-20&101 & 530 & 161 & 94&886 
 \\ 
 \hline 
gpt-4o-2024-08-06&98 & 536 & 151 & 95&880 
 \\ 
 \hline 
gpt-4o-2024-05-13&100 & 534 & 163 & 100&897 
 \\ 
 \hline 
chatgpt-4o-latest&105 & 545 & 163 & 97&910 
 \\ 
 \hline 
gpt-4o-mini-2024-07-18&87 & 476 & 144 & 71&778 
 \\ 
 \hline 
gpt-4-turbo&94 & 514 & 156 & 89&853 
 \\ 
 \hline 
gpt-4-turbo-2024-04-09&97 & 512 & 153 & 89&851 
 \\ 
 \hline 
gemini-3-flash-preview-result&112 & 592 & 210 & 110&1024 
 \\ 
 \hline 
gemini-2.5-pro&107 & 587 & 201 & 110&1005 
 \\ 
 \hline 
gemini-2.5-flash&106&554&176&96&932 
 \\ 
 \hline 
gemini-2.5-flash-lite&101 & 501 & 155 & 82&839 
 \\ 
 \hline 
gemini-2.5-flash-lite-preview-09-2025&98 & 505 & 152 & 84&839 
 \\ 
 \hline 
gemini-2.0-flash&99 & 536 & 151 & 87&873 
 \\ 
 \hline 
gemini-2.0-flash-lite&98 & 514 & 141 & 84&837 
 \\ 
 \hline 

claude-3-haiku-20240307&82 & 423 & 130 & 72&707 
 \\ 
 \hline 
claude-haiku-4-5-20251001&100 & 526 & 151 & 83&860 
 \\ 
 \hline 
claude-opus-4-5-20251101&108 & 583 & 174 & 106&971 
 \\ 
 \hline 

claude-sonnet-4.5-result&104.7&535.0&161.7&98.0&899.3
 \\ 
 \hline 
Maverick&96.7&520.3&150.3&87.0&854.3 
 \\ 
 \hline 
mistral-small3.2&93 & 481 & 126 & 80&780 
 \\ 
 \hline 
mistral-small3.1:latest&97 & 496 & 138 & 81&812 
 \\ 
 \hline 
ministral-3-3b&73 & 399 & 110 & 70&652 
 \\ 
 \hline 
ministral-3-8b&90 & 445 & 122 & 82&739 
 \\ 
 \hline 
ministral-3-14b&90 & 473 & 130 & 85&778 
 \\ 
 \hline 
nvidia-Nemotron-Nano&95 & 492 & 150 & 85&822 
 \\ 
 \hline 
Qwen2.5-VL-72B-Instruct&94 & 535 & 142 & 93&864 
 \\ 
 \hline 
qwen2.5vl:32b&97 & 506 & 135 & 84&822 
 \\ 
 \hline 
qwen2.5vl:3b&85 & 387 & 104 & 65&641 
 \\ 
 \hline 
qwen2.5vl:7b&81 & 446 & 113 & 69&709 
 \\ 
 \hline 
google-gemma-3-27b-it&95 & 508 & 139 & 87&829 
 \\ 
 \hline 
gemma3:latest&76 & 384 & 84 & 63&607 
 \\ 
 \hline 
gemma3:12b&92 & 472 & 125 & 69&758 
 \\ 
 \hline 
granite3.2-vision:2b&61 & 297 & 74 & 53&485 
 \\ 
 \hline 
ibm-granite3.3-vision-2b&68 & 318 & 77 & 57&520 
 \\ 
 \hline 
llava:34b&86 & 436 & 107 & 71&700 
 \\ 
 \hline 
llava:7b&61 & 330 & 80 & 60&531 
 \\ 
 \hline 
bakllava:latest&54 & 274 & 55 & 38&421 
 \\ 
 \hline 
llava-phi3:latest&71 & 382 & 93 & 60&606 
 \\ 
 \hline 
minicpm-v:latest&69 & 379 & 80 & 64&592 
 \\ 
 \hline 
llava-llama3:latest&69 & 296 & 90 & 58&513 
 \\ 
 \hline 
\end{tabular}

\end{table}

\section{Threats to Validity} \label{threats}
This section outlines the threats to validity of our study and how they are addressed.
First, we aim to include a comprehensive list of LLMs in our evaluation. However, the frequent release of new models makes this infeasible. Therefore, we select a representative set of multimodal models, which consists of commercial and open-source models.

Second, as the exams and their solutions are publicly available, it is possible that at least a subset of the studied LLMs saw some questions during training, which would have influenced their performance in those tasks. To mitigate this threat, we introduced RQ3, which transforms the questions using metamorphic relations and evaluates the top performing models.

Third, the study requires correct extractions of questions and answers from the files. To the best of our knowledge, there is no Python library available that can extract information from a PDF file with 100\% accuracy. To mitigate this limitation, we manually verify the extracted information.

Fourth, one of the sample exams provided by ISTQB in the expert category (i.e., CTEL-TM) does not mention marks or question level for questions. 
To maintain consistency in our evaluation, we deduced the marks and corresponding question levels according to the official rules and guidelines provided by ISTQB. This deduction has been verified by the professional holding a core advanced certification from ISTQB.

Finally, we utilize certification exams from one software testing certification body. However, that selected certification body is representative of software certification bodies. Furthermore, in the future, we plan to include certification exams from other fields and other certification bodies. 

\section{Limitations of Language Models}\label{sec:limitation}
Although this study highlights the capabilities of language models in attempting certification exams, there are still some limitations that must be acknowledged. In this section, we discuss the limitations observed of the language models.

Regarding the limitations of language models, one notable limitation is the tendency to hallucinate, where the model generates responses that are irrelevant, inaccurate, or fabricated.
Examples of hallucination include producing strings of characters that do not make any sense, restating the questions, or generating factually incorrect statements about the exam. For instance, one model responded with "ISTQB does not provide multiple-choice questions for certification exams," which is incorrect, as the majority of the exams are entirely based on multiple-choice questions, such as CTFL, CTFL-TM.
Only the expert-level exams include open-ended questions.
Furthermore, some models often select an option that is not available. For instance, when there are four possible options (i.e., `a',`b',`c',`d') and one option should be selected as the answer, the model chose the option `R'.
Moreover, some models produced an option that was not available by combining two options together to be the best possible option. 

Similarly, another limitation of language models that we observed in the study is adherence to instructions. Some models  failed to follow the option selection instructions mentioned in the question.
For example, if two options are to be selected for a question, some models select only one for some questions, causing them to lose marks in the exam. 

In addition, we identify another  limitation of language models in their ability to handle visual content. There are 31 questions that contain at least one image. One of the models, llava-llama-3, provides a description of the image rather than providing an answer to the question.
Furthermore, some models ask a question requesting clarification for the question rather than selecting an option. For example, ``In this image, there is an explanation for coffee dispensers, can you provide a flow chart diagram for these?''

\section{Related Work}\label{sec:relatedworks}
With recent advancements in language models, several studies have been performed to test the capabilities of language models.

Nori et al.\cite{nori2023capabilities} evaluated the capabilities of GPT-4 on on medical competency examinations and benchmark datasets.
Lothritz et al.~\cite{lothritz2025testing} evaluated the capabilities of 53 LLMs for Luxembourgish.
Zhang et al.~\cite{zhang2024benchmarking} performed evaluation of 10 diverse LLMs on news summarization.
Ni et al.~\cite{ni2024l2ceval} proposed L2CEval,which is a systematic evaluation of the language-to-code capabilities (i.e., semantic parsing, math reasoning, and Python programming) of 54 language models.
Tu et al.~\cite{tu2024carl} proposed a dataset, CARL-GT, to evaluate the casual reasoning capabilities of LLMs. Furthermore, they evaluated casual reasoning capabilities of five open-source models.  
Fu et al.~\cite{fu2023large} evaluated three language models in their ability to assess the factual consistency of summaries generated by the large language model. 
Shi et al.~\cite{shi2023chef} proposed ChEF, Comprehensive Evaluation Framework, to evaluate capabilities and
limitations of multimodal large language models. They evaluated on nine language models.
Chen et al.~\cite{chen2025benchmarking} evaluated the capabilities of three language models in completing the homework for an undergraduate-level circuit. analysis course

The closest work to our study is work by Retzlaff et al.~\cite{retzlaff2025automation}. Although this work is available only on the archive, and it has been peer reviewed yet. We still include this in our related work for the sake of completeness.
In this study, they evaluated two language models GPT-4o and GEMINI 2.0 Flash on only four certification exams (A4Q Software Development Engineer in Test, ISTQB Certified Tester Foundation Level, Advanced Level Test Manager, and Expert Level Test Manager). Their results indicate that both models pass all of the exams.
Our study provides a more comprehensive comparison of 60 language models on a wider range of certification exams.
Furthermore, we evaluate the robustness of best performing language models by querying them with metamorphically transformed exam questions.
In addition to that, we also provide a leaderboard and complete replication package to extract, verify the questions, and query the language model.


\section{Conclusion}\label{sec:conclusion}
In this study, we investigate the capabilities of a large number of language models in attempting the certification exams. 
We evaluated 60 language models, from commercial vendors and open source community, of various sizes, on 30 different ISTQB software testing certification exams, resulting in 1171 questions for every model. In total, we executed $\sim$71k prompts. 

The results indicate that Gemini-3-flash-preview and gpt-5-2025-08-07 score at least 65\% on all of the certification exams. 
We further compared different language models based on their ability to answer questions with different question levels. Gemini-3-flash-preview outperform other models on all the question levels.
In addition to these, we also compared models based on their sizes. 
Furthermore, we also compared only the open-source models. In open-source models, Qwen2.5 variants performs better than the other models for question level K1, K2, and K4, while Llama 4 Maverick performs better on questions with level K3.

In order to ensure that models demonstrate a robust understanding of the concepts of software testing rather than relying on memorization,
we transformed the exam questions using metamorphic relations and evaluated the top performance models. The results show a drop in performance, indicating that even high-performing models can be influenced by small changes in questions, highlighting the need for further research on improving the robustness of language models.

We further investigated the questions that none of the language models answered correctly and identified reasons for these failures. First, the question does not contain enough information to make a well-informed decision about the answer. Second, we found that the options are too similar to judge properly.
Based on the findings, we recommend that the questions in the exams should contain enough information for a clear and justifiable choice. Furthermore, the distractors should be conceptually different than the correct one rather than providing equally plausible options in the same context 

In future work, we plan to include a broader and more diverse range of large language models. Additionally, we plan to  consider a wider set of certification domains beyond ISTQB, such as cybersecurity, and project management.
Furthermore, we intend to investigate the ability of language models to generate certification exams following the guidelines provided by certification bodies.

\bibliographystyle{ACM-Reference-Format}
\bibliography{refs}


\begin{thebibliography}{60}


\ifx \showCODEN    \undefined \def \showCODEN     #1{\unskip}     \fi
\ifx \showISBNx    \undefined \def \showISBNx     #1{\unskip}     \fi
\ifx \showISBNxiii \undefined \def \showISBNxiii  #1{\unskip}     \fi
\ifx \showISSN     \undefined \def \showISSN      #1{\unskip}     \fi
\ifx \showLCCN     \undefined \def \showLCCN      #1{\unskip}     \fi
\ifx \shownote     \undefined \def \shownote      #1{#1}          \fi
\ifx \showarticletitle \undefined \def \showarticletitle #1{#1}   \fi
\ifx \showURL      \undefined \def \showURL       {\relax}        \fi
\providecommand\bibfield[2]{#2}
\providecommand\bibinfo[2]{#2}
\providecommand\natexlab[1]{#1}
\providecommand\showeprint[2][]{arXiv:#2}

\bibitem[Anon(2026)]%
        {supp}
\bibfield{author}{\bibinfo{person}{Anon}.} \bibinfo{year}{2026}\natexlab{}.
\newblock \bibinfo{title}{Replication Package For "Can Language Models Pass Software Testing Certification Exams: a case study"}.
\newblock
\urldef\tempurl%
\url{will-be-released-after-publication}
\showURL{%
\tempurl}


\bibitem[Bai et~al\mbox{.}(2025)]%
        {bai2025qwen2}
\bibfield{author}{\bibinfo{person}{Shuai Bai}, \bibinfo{person}{Keqin Chen}, \bibinfo{person}{Xuejing Liu}, \bibinfo{person}{Jialin Wang}, \bibinfo{person}{Wenbin Ge}, \bibinfo{person}{Sibo Song}, \bibinfo{person}{Kai Dang}, \bibinfo{person}{Peng Wang}, \bibinfo{person}{Shijie Wang}, \bibinfo{person}{Jun Tang}, {et~al\mbox{.}}} \bibinfo{year}{2025}\natexlab{}.
\newblock \showarticletitle{Qwen2. 5-vl technical report}.
\newblock \bibinfo{journal}{\emph{arXiv preprint arXiv:2502.13923}} (\bibinfo{year}{2025}).
\newblock


\bibitem[Basant et~al\mbox{.}(2025)]%
        {basant2025nvidia}
\bibfield{author}{\bibinfo{person}{Aarti Basant}, \bibinfo{person}{Abhijit Khairnar}, \bibinfo{person}{Abhijit Paithankar}, \bibinfo{person}{Abhinav Khattar}, \bibinfo{person}{Adithya Renduchintala}, \bibinfo{person}{Aditya Malte}, \bibinfo{person}{Akhiad Bercovich}, \bibinfo{person}{Akshay Hazare}, \bibinfo{person}{Alejandra Rico}, \bibinfo{person}{Aleksander Ficek}, {et~al\mbox{.}}} \bibinfo{year}{2025}\natexlab{}.
\newblock \showarticletitle{Nvidia nemotron nano 2: An accurate and efficient hybrid mamba-transformer reasoning model}.
\newblock \bibinfo{journal}{\emph{arXiv preprint arXiv:2508.14444}} (\bibinfo{year}{2025}).
\newblock


\bibitem[Chen et~al\mbox{.}(2025)]%
        {chen2025benchmarking}
\bibfield{author}{\bibinfo{person}{Liangliang Chen}, \bibinfo{person}{Zhihao Qin}, \bibinfo{person}{Yiming Guo}, \bibinfo{person}{Jacqueline Rohde}, {and} \bibinfo{person}{Ying Zhang}.} \bibinfo{year}{2025}\natexlab{}.
\newblock \showarticletitle{Benchmarking large language models on homework assessment in circuit analysis}.
\newblock \bibinfo{journal}{\emph{International Journal of Artificial Intelligence in Education}} (\bibinfo{year}{2025}), \bibinfo{pages}{1--62}.
\newblock


\bibitem[Cho et~al\mbox{.}(2025)]%
        {cho2025metamorphic}
\bibfield{author}{\bibinfo{person}{Steven Cho}, \bibinfo{person}{Stefano Ruberto}, {and} \bibinfo{person}{Valerio Terragni}.} \bibinfo{year}{2025}\natexlab{}.
\newblock \showarticletitle{Metamorphic testing of large language models for natural language processing}. In \bibinfo{booktitle}{\emph{2025 IEEE International Conference on Software Maintenance and Evolution (ICSME)}}. IEEE, \bibinfo{pages}{174--186}.
\newblock


\bibitem[Comanici et~al\mbox{.}(2025)]%
        {comanici2025gemini}
\bibfield{author}{\bibinfo{person}{Gheorghe Comanici}, \bibinfo{person}{Eric Bieber}, \bibinfo{person}{Mike Schaekermann}, \bibinfo{person}{Ice Pasupat}, \bibinfo{person}{Noveen Sachdeva}, \bibinfo{person}{Inderjit Dhillon}, \bibinfo{person}{Marcel Blistein}, \bibinfo{person}{Ori Ram}, \bibinfo{person}{Dan Zhang}, \bibinfo{person}{Evan Rosen}, {et~al\mbox{.}}} \bibinfo{year}{2025}\natexlab{}.
\newblock \showarticletitle{Gemini 2.5: Pushing the frontier with advanced reasoning, multimodality, long context, and next generation agentic capabilities}.
\newblock \bibinfo{journal}{\emph{arXiv preprint arXiv:2507.06261}} (\bibinfo{year}{2025}).
\newblock


\bibitem[Contributors(2023)]%
        {2023xtuner}
\bibfield{author}{\bibinfo{person}{XTuner Contributors}.} \bibinfo{year}{2023}\natexlab{}.
\newblock \bibinfo{title}{XTuner: A Toolkit for Efficiently Fine-tuning LLM}.
\newblock \bibinfo{howpublished}{\url{https://github.com/InternLM/xtuner}}.
\newblock


\bibitem[Dakhel et~al\mbox{.}(2024)]%
        {dakhel2024effective}
\bibfield{author}{\bibinfo{person}{Arghavan~Moradi Dakhel}, \bibinfo{person}{Amin Nikanjam}, \bibinfo{person}{Vahid Majdinasab}, \bibinfo{person}{Foutse Khomh}, {and} \bibinfo{person}{Michel~C Desmarais}.} \bibinfo{year}{2024}\natexlab{}.
\newblock \showarticletitle{Effective test generation using pre-trained large language models and mutation testing}.
\newblock \bibinfo{journal}{\emph{Information and Software Technology}}  \bibinfo{volume}{171} (\bibinfo{year}{2024}), \bibinfo{pages}{107468}.
\newblock


\bibitem[Fatima et~al\mbox{.}(2024)]%
        {fatima2024flakyfix}
\bibfield{author}{\bibinfo{person}{Sakina Fatima}, \bibinfo{person}{Hadi Hemmati}, {and} \bibinfo{person}{Lionel~C Briand}.} \bibinfo{year}{2024}\natexlab{}.
\newblock \showarticletitle{FlakyFix: Using large language models for predicting flaky test fix categories and test code repair}.
\newblock \bibinfo{journal}{\emph{IEEE Transactions on Software Engineering}} \bibinfo{volume}{50}, \bibinfo{number}{12} (\bibinfo{year}{2024}), \bibinfo{pages}{3146--3171}.
\newblock


\bibitem[Fu et~al\mbox{.}(2023)]%
        {fu2023large}
\bibfield{author}{\bibinfo{person}{Xue-Yong Fu}, \bibinfo{person}{Md~Tahmid~Rahman Laskar}, \bibinfo{person}{Cheng Chen}, {and} \bibinfo{person}{Shashi~Bhushan Tn}.} \bibinfo{year}{2023}\natexlab{}.
\newblock \showarticletitle{Are large language models reliable judges? a study on the factuality evaluation capabilities of LLMs}. In \bibinfo{booktitle}{\emph{Proceedings of the third workshop on natural language generation, evaluation, and metrics (gem)}}. \bibinfo{pages}{310--316}.
\newblock


\bibitem[{Google}(2025)]%
        {Gemini30}
\bibfield{author}{\bibinfo{person}{{Google}}.} \bibinfo{year}{2025}\natexlab{}.
\newblock \bibinfo{title}{Gemini 3.0}.
\newblock \bibinfo{howpublished}{\url{https://aistudio.google.com/models/gemini-3}}.
\newblock


\bibitem[Gu et~al\mbox{.}(2024)]%
        {gu2024testart}
\bibfield{author}{\bibinfo{person}{Siqi Gu}, \bibinfo{person}{Quanjun Zhang}, \bibinfo{person}{Kecheng Li}, \bibinfo{person}{Chunrong Fang}, \bibinfo{person}{Fangyuan Tian}, \bibinfo{person}{Liuchuan Zhu}, \bibinfo{person}{Jianyi Zhou}, {and} \bibinfo{person}{Zhenyu Chen}.} \bibinfo{year}{2024}\natexlab{}.
\newblock \showarticletitle{Testart: Improving llm-based unit testing via co-evolution of automated generation and repair iteration}.
\newblock \bibinfo{journal}{\emph{arXiv preprint arXiv:2408.03095}} (\bibinfo{year}{2024}).
\newblock


\bibitem[Guilherme and Vincenzi(2023)]%
        {guilherme2023initial}
\bibfield{author}{\bibinfo{person}{Vitor Guilherme} {and} \bibinfo{person}{Auri Vincenzi}.} \bibinfo{year}{2023}\natexlab{}.
\newblock \showarticletitle{An initial investigation of ChatGPT unit test generation capability}. In \bibinfo{booktitle}{\emph{Proceedings of the 8th Brazilian Symposium on Systematic and Automated Software Testing}}. \bibinfo{pages}{15--24}.
\newblock


\bibitem[Hashtroudi et~al\mbox{.}(2023)]%
        {hashtroudi2023automated}
\bibfield{author}{\bibinfo{person}{Sepehr Hashtroudi}, \bibinfo{person}{Jiho Shin}, \bibinfo{person}{Hadi Hemmati}, {and} \bibinfo{person}{Song Wang}.} \bibinfo{year}{2023}\natexlab{}.
\newblock \showarticletitle{Automated test case generation using code models and domain adaptation}.
\newblock \bibinfo{journal}{\emph{CoRR}} (\bibinfo{year}{2023}).
\newblock


\bibitem[Hossain and Dwyer(2025)]%
        {HossainDwyer2024_TOGLL}
\bibfield{author}{\bibinfo{person}{Soneya~Binta Hossain} {and} \bibinfo{person}{Matthew~B. Dwyer}.} \bibinfo{year}{2025}\natexlab{}.
\newblock \showarticletitle{TOGLL: Correct and Strong Test Oracle Generation with LLMS}. In \bibinfo{booktitle}{\emph{2025 IEEE/ACM 47th International Conference on Software Engineering (ICSE)}}. \bibinfo{pages}{1475--1487}.
\newblock
\href{https://doi.org/10.1109/ICSE55347.2025.00098}{doi:\nolinkurl{10.1109/ICSE55347.2025.00098}}


\bibitem[Hyun et~al\mbox{.}(2024)]%
        {hyun2024metal}
\bibfield{author}{\bibinfo{person}{Sangwon Hyun}, \bibinfo{person}{Mingyu Guo}, {and} \bibinfo{person}{M~Ali Babar}.} \bibinfo{year}{2024}\natexlab{}.
\newblock \showarticletitle{METAL: metamorphic testing framework for analyzing large-language model qualities}. In \bibinfo{booktitle}{\emph{2024 IEEE Conference on Software Testing, Verification and Validation (ICST)}}. IEEE, \bibinfo{pages}{117--128}.
\newblock


\bibitem[Khandaker et~al\mbox{.}(2025)]%
        {KhandakerKifetewPrandiSusi2025_AugmenTest}
\bibfield{author}{\bibinfo{person}{Shaker~Mahmud Khandaker}, \bibinfo{person}{Fitsum~Meshesha Kifetew}, \bibinfo{person}{Davide Prandi}, {and} \bibinfo{person}{Angelo Susi}.} \bibinfo{year}{2025}\natexlab{}.
\newblock \showarticletitle{AugmenTest: Enhancing Tests with LLM-Driven Oracles}. In \bibinfo{booktitle}{\emph{Proceedings of the IEEE International Conference on Software Testing, Verification and Validation (ICST)}}. \bibinfo{publisher}{IEEE}, \bibinfo{pages}{279--289}.
\newblock
\href{https://doi.org/10.1109/ICST62969.2025.10988926}{doi:\nolinkurl{10.1109/ICST62969.2025.10988926}}


\bibitem[Le et~al\mbox{.}(2025)]%
        {le2025testweaver}
\bibfield{author}{\bibinfo{person}{Cuong~Chi Le}, \bibinfo{person}{Cuong~Duc Van}, \bibinfo{person}{Tung~Duy Vu}, \bibinfo{person}{Thai Minh~Pham Vu}, \bibinfo{person}{Hoang~Nhat Phan}, \bibinfo{person}{Huy~Nhat Phan}, {and} \bibinfo{person}{Tien~N Nguyen}.} \bibinfo{year}{2025}\natexlab{}.
\newblock \showarticletitle{TestWeaver: Execution-aware, Feedback-driven Regression Testing Generation with Large Language Models}.
\newblock \bibinfo{journal}{\emph{arXiv preprint arXiv:2508.01255}} (\bibinfo{year}{2025}).
\newblock


\bibitem[Li et~al\mbox{.}(2024)]%
        {li2024empirical}
\bibfield{author}{\bibinfo{person}{Zongjie Li}, \bibinfo{person}{Wenying Qiu}, \bibinfo{person}{Pingchuan Ma}, \bibinfo{person}{Yichen Li}, \bibinfo{person}{You Li}, \bibinfo{person}{Sijia He}, \bibinfo{person}{Baozheng Jiang}, \bibinfo{person}{Shuai Wang}, {and} \bibinfo{person}{Weixi Gu}.} \bibinfo{year}{2024}\natexlab{}.
\newblock \showarticletitle{An empirical study on large language models in accuracy and robustness under chinese industrial scenarios}.
\newblock \bibinfo{journal}{\emph{arXiv preprint arXiv:2402.01723}} (\bibinfo{year}{2024}).
\newblock


\bibitem[Liu et~al\mbox{.}(2026)]%
        {liu2026ministral}
\bibfield{author}{\bibinfo{person}{Alexander~H Liu}, \bibinfo{person}{Kartik Khandelwal}, \bibinfo{person}{Sandeep Subramanian}, \bibinfo{person}{Victor Jouault}, \bibinfo{person}{Abhinav Rastogi}, \bibinfo{person}{Adrien Sad{\'e}}, \bibinfo{person}{Alan Jeffares}, \bibinfo{person}{Albert Jiang}, \bibinfo{person}{Alexandre Cahill}, \bibinfo{person}{Alexandre Gavaudan}, {et~al\mbox{.}}} \bibinfo{year}{2026}\natexlab{}.
\newblock \showarticletitle{Ministral 3}.
\newblock \bibinfo{journal}{\emph{arXiv preprint arXiv:2601.08584}} (\bibinfo{year}{2026}).
\newblock


\bibitem[Liu et~al\mbox{.}(2024a)]%
        {liu2024improved}
\bibfield{author}{\bibinfo{person}{Haotian Liu}, \bibinfo{person}{Chunyuan Li}, \bibinfo{person}{Yuheng Li}, {and} \bibinfo{person}{Yong~Jae Lee}.} \bibinfo{year}{2024}\natexlab{a}.
\newblock \showarticletitle{Improved baselines with visual instruction tuning}. In \bibinfo{booktitle}{\emph{Proceedings of the IEEE/CVF conference on computer vision and pattern recognition}}. \bibinfo{pages}{26296--26306}.
\newblock


\bibitem[Liu et~al\mbox{.}(2024b)]%
        {liu2024fix}
\bibfield{author}{\bibinfo{person}{Jun Liu}, \bibinfo{person}{Jiwei Yan}, \bibinfo{person}{Yuanyuan Xie}, \bibinfo{person}{Jun Yan}, {and} \bibinfo{person}{Jian Zhang}.} \bibinfo{year}{2024}\natexlab{b}.
\newblock \showarticletitle{Fix the Tests: Augmenting LLMs to Repair Test Cases with Static Collector and Neural Reranker}. In \bibinfo{booktitle}{\emph{2024 IEEE 35th International Symposium on Software Reliability Engineering (ISSRE)}}. IEEE, \bibinfo{pages}{367--378}.
\newblock


\bibitem[Lothritz and Cabot(2025)]%
        {lothritz2025testing}
\bibfield{author}{\bibinfo{person}{Cedric Lothritz} {and} \bibinfo{person}{Jordi Cabot}.} \bibinfo{year}{2025}\natexlab{}.
\newblock \showarticletitle{Testing Low-Resource Language Support in LLMs Using Language Proficiency Exams: the Case of Luxembourgish}.
\newblock \bibinfo{journal}{\emph{arXiv preprint arXiv:2504.01667}} (\bibinfo{year}{2025}).
\newblock


\bibitem[{Mistral}(2025a)]%
        {mistralsmall3.1}
\bibfield{author}{\bibinfo{person}{{Mistral}}.} \bibinfo{year}{2025}\natexlab{a}.
\newblock \bibinfo{title}{Mistral Small 3.1}.
\newblock \bibinfo{howpublished}{\url{https://mistral.ai/news/mistral-small-3-1}}.
\newblock


\bibitem[{Mistral}(2025b)]%
        {mistralsmall3.2}
\bibfield{author}{\bibinfo{person}{{Mistral}}.} \bibinfo{year}{2025}\natexlab{b}.
\newblock \bibinfo{title}{Mistral Small 3.2}.
\newblock \bibinfo{howpublished}{\url{https://docs.mistral.ai/models/mistral-small-3-2-25-06}}.
\newblock


\bibitem[Molina et~al\mbox{.}(2025)]%
        {molina2025test}
\bibfield{author}{\bibinfo{person}{Facundo Molina}, \bibinfo{person}{Alessandra Gorla}, {and} \bibinfo{person}{Marcelo d’Amorim}.} \bibinfo{year}{2025}\natexlab{}.
\newblock \showarticletitle{Test Oracle Automation in the era of LLMs}.
\newblock \bibinfo{journal}{\emph{ACM Transactions on Software Engineering and Methodology}} \bibinfo{volume}{34}, \bibinfo{number}{5} (\bibinfo{year}{2025}), \bibinfo{pages}{1--24}.
\newblock


\bibitem[{Nebius}(2025)]%
        {nebius_token_factory}
\bibfield{author}{\bibinfo{person}{{Nebius}}.} \bibinfo{year}{2025}\natexlab{}.
\newblock \bibinfo{title}{{Nebius Token Factory}}.
\newblock \bibinfo{howpublished}{\url{https://tokenfactory.nebius.com/}}.
\newblock
\urldef\tempurl%
\url{https://tokenfactory.nebius.com/}
\showURL{%
\tempurl}
\newblock
\shownote{Production-grade AI inference platform supporting 60+ open-source models, with OpenAI-compatible API for scalable deployment and inference.}.


\bibitem[Ni et~al\mbox{.}(2024)]%
        {ni2024l2ceval}
\bibfield{author}{\bibinfo{person}{Ansong Ni}, \bibinfo{person}{Pengcheng Yin}, \bibinfo{person}{Yilun Zhao}, \bibinfo{person}{Martin Riddell}, \bibinfo{person}{Troy Feng}, \bibinfo{person}{Rui Shen}, \bibinfo{person}{Stephen Yin}, \bibinfo{person}{Ye Liu}, \bibinfo{person}{Semih Yavuz}, \bibinfo{person}{Caiming Xiong}, {et~al\mbox{.}}} \bibinfo{year}{2024}\natexlab{}.
\newblock \showarticletitle{L2ceval: Evaluating language-to-code generation capabilities of large language models}.
\newblock \bibinfo{journal}{\emph{Transactions of the Association for Computational Linguistics}}  \bibinfo{volume}{12} (\bibinfo{year}{2024}), \bibinfo{pages}{1311--1329}.
\newblock


\bibitem[Nori et~al\mbox{.}(2023)]%
        {nori2023capabilities}
\bibfield{author}{\bibinfo{person}{Harsha Nori}, \bibinfo{person}{Nicholas King}, \bibinfo{person}{Scott~Mayer McKinney}, \bibinfo{person}{Dean Carignan}, {and} \bibinfo{person}{Eric Horvitz}.} \bibinfo{year}{2023}\natexlab{}.
\newblock \showarticletitle{Capabilities of gpt-4 on medical challenge problems}.
\newblock \bibinfo{journal}{\emph{arXiv preprint arXiv:2303.13375}} (\bibinfo{year}{2023}).
\newblock


\bibitem[{OpenAI}(2023a)]%
        {openai2025gpt4}
\bibfield{author}{\bibinfo{person}{{OpenAI}}.} \bibinfo{year}{2023}\natexlab{a}.
\newblock \bibinfo{title}{GPT-4}.
\newblock \bibinfo{howpublished}{\url{https://platform.openai.com/docs/models/gpt-4}}.
\newblock


\bibitem[{OpenAI}(2023b)]%
        {openai2025gpt4turbo}
\bibfield{author}{\bibinfo{person}{{OpenAI}}.} \bibinfo{year}{2023}\natexlab{b}.
\newblock \bibinfo{title}{GPT-4 Turbo}.
\newblock \bibinfo{howpublished}{\url{https://platform.openai.com/docs/models/gpt-4-turbo}}.
\newblock


\bibitem[{OpenAI}(2024a)]%
        {openai20254o}
\bibfield{author}{\bibinfo{person}{{OpenAI}}.} \bibinfo{year}{2024}\natexlab{a}.
\newblock \bibinfo{title}{ChatGPT-4o-latest}.
\newblock \bibinfo{howpublished}{\url{ https://platform.openai.com/docs/models/chatgpt-4o-latest}}.
\newblock


\bibitem[{OpenAI}(2024b)]%
        {openai2025o4o}
\bibfield{author}{\bibinfo{person}{{OpenAI}}.} \bibinfo{year}{2024}\natexlab{b}.
\newblock \bibinfo{title}{GPT-4o}.
\newblock \bibinfo{howpublished}{\url{ https://platform.openai.com/docs/models/gpt-4o}}.
\newblock


\bibitem[{OpenAI}(2024c)]%
        {openai2025o3}
\bibfield{author}{\bibinfo{person}{{OpenAI}}.} \bibinfo{year}{2024}\natexlab{c}.
\newblock \bibinfo{title}{O3}.
\newblock \bibinfo{howpublished}{\url{ https://platform.openai.com/docs/models/o3}}.
\newblock


\bibitem[{OpenAI}(2024d)]%
        {openai2025o4}
\bibfield{author}{\bibinfo{person}{{OpenAI}}.} \bibinfo{year}{2024}\natexlab{d}.
\newblock \bibinfo{title}{o4-mini}.
\newblock \bibinfo{howpublished}{\url{ https://platform.openai.com/docs/models/o4-mini}}.
\newblock


\bibitem[{OpenAI}(2025a)]%
        {openai2025gpt41}
\bibfield{author}{\bibinfo{person}{{OpenAI}}.} \bibinfo{year}{2025}\natexlab{a}.
\newblock \bibinfo{title}{GPT-4.1}.
\newblock \bibinfo{howpublished}{\url{https://platform.openai.com/docs/models/gpt-4.1}}.
\newblock


\bibitem[{OpenAI}(2025b)]%
        {openai2025gpt41mini}
\bibfield{author}{\bibinfo{person}{{OpenAI}}.} \bibinfo{year}{2025}\natexlab{b}.
\newblock \bibinfo{title}{GPT-4.1 Mini}.
\newblock \bibinfo{howpublished}{\url{https://platform.openai.com/docs/models/gpt-4.1-mini}}.
\newblock


\bibitem[{OpenAI}(2025c)]%
        {openai2025gpt41nano}
\bibfield{author}{\bibinfo{person}{{OpenAI}}.} \bibinfo{year}{2025}\natexlab{c}.
\newblock \bibinfo{title}{GPT-4.1 Nano}.
\newblock \bibinfo{howpublished}{\url{https://platform.openai.com/docs/models/gpt-4.1-nano}}.
\newblock


\bibitem[{OpenAI}(2025d)]%
        {openai2025gpt5}
\bibfield{author}{\bibinfo{person}{{OpenAI}}.} \bibinfo{year}{2025}\natexlab{d}.
\newblock \bibinfo{title}{GPT-5}.
\newblock \bibinfo{howpublished}{\url{https://platform.openai.com/docs/models/gpt-5}}.
\newblock


\bibitem[{OpenAI}(2025e)]%
        {openai2025gpt5mini}
\bibfield{author}{\bibinfo{person}{{OpenAI}}.} \bibinfo{year}{2025}\natexlab{e}.
\newblock \bibinfo{title}{GPT-5 Mini}.
\newblock \bibinfo{howpublished}{\url{https://platform.openai.com/docs/models/gpt-5-mini}}.
\newblock


\bibitem[{OpenAI}(2025f)]%
        {openai2025gpt5nano}
\bibfield{author}{\bibinfo{person}{{OpenAI}}.} \bibinfo{year}{2025}\natexlab{f}.
\newblock \bibinfo{title}{GPT-5 Nano}.
\newblock \bibinfo{howpublished}{\url{https://platform.openai.com/docs/models/gpt-5-nano}}.
\newblock


\bibitem[{OpenAI}(2025g)]%
        {openai2025gpt51}
\bibfield{author}{\bibinfo{person}{{OpenAI}}.} \bibinfo{year}{2025}\natexlab{g}.
\newblock \bibinfo{title}{GPT-5.1}.
\newblock \bibinfo{howpublished}{\url{https://openai.com/index/introducing-gpt-5-1/}}.
\newblock


\bibitem[{OpenAI}(2025h)]%
        {openai2025gpt52}
\bibfield{author}{\bibinfo{person}{{OpenAI}}.} \bibinfo{year}{2025}\natexlab{h}.
\newblock \bibinfo{title}{GPT-5.2}.
\newblock \bibinfo{howpublished}{\url{https://openai.com/index/introducing-gpt-5-2/}}.
\newblock


\bibitem[{OpenAI}(2025i)]%
        {openaiChangenotes}
\bibfield{author}{\bibinfo{person}{{OpenAI}}.} \bibinfo{year}{2025}\natexlab{i}.
\newblock \bibinfo{title}{OpenAI Release notes}.
\newblock \bibinfo{howpublished}{\url{help.openai.com/en/articles/6825453-chatgpt-release-notes}}.
\newblock


\bibitem[Pan et~al\mbox{.}(2024)]%
        {PanGhalebBriand2024_LTM}
\bibfield{author}{\bibinfo{person}{Rongqi Pan}, \bibinfo{person}{Taher~A. Ghaleb}, {and} \bibinfo{person}{Lionel~C. Briand}.} \bibinfo{year}{2024}\natexlab{}.
\newblock \showarticletitle{LTM: Scalable and Black-Box Similarity-Based Test Suite Minimization Based on Language Models}.
\newblock \bibinfo{journal}{\emph{IEEE Transactions on Software Engineering}} \bibinfo{volume}{50}, \bibinfo{number}{4} (\bibinfo{year}{2024}), \bibinfo{pages}{3053--3070}.
\newblock
\href{https://doi.org/10.1109/TSE.2024.3469582}{doi:\nolinkurl{10.1109/TSE.2024.3469582}}


\bibitem[Retzlaff(2025)]%
        {retzlaff2025automation}
\bibfield{author}{\bibinfo{person}{Niklas Retzlaff}.} \bibinfo{year}{2025}\natexlab{}.
\newblock \showarticletitle{From Automation to Certification: Benchmarking AI Chatbots in Software Testing}.
\newblock  (\bibinfo{year}{2025}).
\newblock


\bibitem[Sch{\"a}fer et~al\mbox{.}(2023)]%
        {schafer2023empirical}
\bibfield{author}{\bibinfo{person}{Max Sch{\"a}fer}, \bibinfo{person}{Sarah Nadi}, \bibinfo{person}{Aryaz Eghbali}, {and} \bibinfo{person}{Frank Tip}.} \bibinfo{year}{2023}\natexlab{}.
\newblock \showarticletitle{An empirical evaluation of using large language models for automated unit test generation}.
\newblock \bibinfo{journal}{\emph{IEEE Transactions on Software Engineering}} \bibinfo{volume}{50}, \bibinfo{number}{1} (\bibinfo{year}{2023}), \bibinfo{pages}{85--105}.
\newblock


\bibitem[Shi et~al\mbox{.}(2023)]%
        {shi2023chef}
\bibfield{author}{\bibinfo{person}{Zhelun Shi}, \bibinfo{person}{Zhipin Wang}, \bibinfo{person}{Hongxing Fan}, \bibinfo{person}{Zhenfei Yin}, \bibinfo{person}{Lu Sheng}, \bibinfo{person}{Yu Qiao}, {and} \bibinfo{person}{Jing Shao}.} \bibinfo{year}{2023}\natexlab{}.
\newblock \showarticletitle{Chef: A comprehensive evaluation framework for standardized assessment of multimodal large language models}.
\newblock \bibinfo{journal}{\emph{arXiv preprint arXiv:2311.02692}} (\bibinfo{year}{2023}).
\newblock


\bibitem[Srinivasan et~al\mbox{.}(2018)]%
        {srinivasan2018quality}
\bibfield{author}{\bibinfo{person}{Madhusudan Srinivasan}, \bibinfo{person}{Morteza~Pourreza Shahri}, \bibinfo{person}{Indika Kahanda}, {and} \bibinfo{person}{Upulee Kanewala}.} \bibinfo{year}{2018}\natexlab{}.
\newblock \showarticletitle{Quality assurance of bioinformatics software: a case study of testing a biomedical text processing tool using metamorphic testing}. In \bibinfo{booktitle}{\emph{Proceedings of the 3rd International Workshop on Metamorphic Testing}}. \bibinfo{pages}{26--33}.
\newblock


\bibitem[Taherkhani and Hemmati(2024)]%
        {TaherkhaniHemmati2024_VALTEST}
\bibfield{author}{\bibinfo{person}{Hamed Taherkhani} {and} \bibinfo{person}{Hadi Hemmati}.} \bibinfo{year}{2024}\natexlab{}.
\newblock \showarticletitle{VALTEST: Automated Validation of Language Model Generated Test Cases}.
\newblock \bibinfo{journal}{\emph{arXiv preprint}}  \bibinfo{volume}{abs/2411.08254} (\bibinfo{year}{2024}).
\newblock
\urldef\tempurl%
\url{https://arxiv.org/abs/2411.08254}
\showURL{%
\tempurl}


\bibitem[Tang et~al\mbox{.}(2024)]%
        {tang2024chatGPT}
\bibfield{author}{\bibinfo{person}{Yutian Tang}, \bibinfo{person}{Zhijie Liu}, \bibinfo{person}{Zhichao Zhou}, {and} \bibinfo{person}{Xiapu Luo}.} \bibinfo{year}{2024}\natexlab{}.
\newblock \showarticletitle{Chatgpt vs sbst: A comparative assessment of unit test suite generation}.
\newblock \bibinfo{journal}{\emph{IEEE Transactions on Software Engineering}} \bibinfo{volume}{50}, \bibinfo{number}{6} (\bibinfo{year}{2024}), \bibinfo{pages}{1340--1359}.
\newblock


\bibitem[Team et~al\mbox{.}(2025a)]%
        {team2025gemma}
\bibfield{author}{\bibinfo{person}{Gemma Team}, \bibinfo{person}{Aishwarya Kamath}, \bibinfo{person}{Johan Ferret}, \bibinfo{person}{Shreya Pathak}, \bibinfo{person}{Nino Vieillard}, \bibinfo{person}{Ramona Merhej}, \bibinfo{person}{Sarah Perrin}, \bibinfo{person}{Tatiana Matejovicova}, \bibinfo{person}{Alexandre Ram{\'e}}, \bibinfo{person}{Morgane Rivi{\`e}re}, {et~al\mbox{.}}} \bibinfo{year}{2025}\natexlab{a}.
\newblock \showarticletitle{Gemma 3 technical report}.
\newblock \bibinfo{journal}{\emph{arXiv preprint arXiv:2503.19786}} (\bibinfo{year}{2025}).
\newblock


\bibitem[Team et~al\mbox{.}(2025b)]%
        {team2025granite}
\bibfield{author}{\bibinfo{person}{Granite~Vision Team}, \bibinfo{person}{Leonid Karlinsky}, \bibinfo{person}{Assaf Arbelle}, \bibinfo{person}{Abraham Daniels}, \bibinfo{person}{Ahmed Nassar}, \bibinfo{person}{Amit Alfassi}, \bibinfo{person}{Bo Wu}, \bibinfo{person}{Eli Schwartz}, \bibinfo{person}{Dhiraj Joshi}, \bibinfo{person}{Jovana Kondic}, {et~al\mbox{.}}} \bibinfo{year}{2025}\natexlab{b}.
\newblock \showarticletitle{Granite Vision: a lightweight, open-source multimodal model for enterprise Intelligence}.
\newblock \bibinfo{journal}{\emph{arXiv preprint arXiv:2502.09927}} (\bibinfo{year}{2025}).
\newblock


\bibitem[Tu et~al\mbox{.}(2024)]%
        {tu2024carl}
\bibfield{author}{\bibinfo{person}{Ruibo Tu}, \bibinfo{person}{Hedvig Kjellstr{\"o}m}, \bibinfo{person}{Gustav~Eje Henter}, {and} \bibinfo{person}{Cheng Zhang}.} \bibinfo{year}{2024}\natexlab{}.
\newblock \showarticletitle{CARL-GT: Evaluating Causal Reasoning Capabilities of Large Language Models}.
\newblock \bibinfo{journal}{\emph{arXiv preprint arXiv:2412.17970}} (\bibinfo{year}{2024}).
\newblock


\bibitem[Wang et~al\mbox{.}(2023)]%
        {wang2023mttm}
\bibfield{author}{\bibinfo{person}{Wenxuan Wang}, \bibinfo{person}{Jen-tse Huang}, \bibinfo{person}{Weibin Wu}, \bibinfo{person}{Jianping Zhang}, \bibinfo{person}{Yizhan Huang}, \bibinfo{person}{Shuqing Li}, \bibinfo{person}{Pinjia He}, {and} \bibinfo{person}{Michael~R Lyu}.} \bibinfo{year}{2023}\natexlab{}.
\newblock \showarticletitle{Mttm: Metamorphic testing for textual content moderation software}. In \bibinfo{booktitle}{\emph{2023 IEEE/ACM 45th International Conference on Software Engineering (ICSE)}}. IEEE, \bibinfo{pages}{2387--2399}.
\newblock


\bibitem[Yao et~al\mbox{.}(2024)]%
        {yao2024minicpm}
\bibfield{author}{\bibinfo{person}{Yuan Yao}, \bibinfo{person}{Tianyu Yu}, \bibinfo{person}{Ao Zhang}, \bibinfo{person}{Chongyi Wang}, \bibinfo{person}{Junbo Cui}, \bibinfo{person}{Hongji Zhu}, \bibinfo{person}{Tianchi Cai}, \bibinfo{person}{Haoyu Li}, \bibinfo{person}{Weilin Zhao}, \bibinfo{person}{Zhihui He}, {et~al\mbox{.}}} \bibinfo{year}{2024}\natexlab{}.
\newblock \showarticletitle{MiniCPM-V: A GPT-4V Level MLLM on Your Phone}.
\newblock \bibinfo{journal}{\emph{arXiv preprint arXiv:2408.01800}} (\bibinfo{year}{2024}).
\newblock


\bibitem[Yaraghi et~al\mbox{.}(2025)]%
        {yaraghi2025automated}
\bibfield{author}{\bibinfo{person}{Ahmadreza~Saboor Yaraghi}, \bibinfo{person}{Darren Holden}, \bibinfo{person}{Nafiseh Kahani}, {and} \bibinfo{person}{Lionel Briand}.} \bibinfo{year}{2025}\natexlab{}.
\newblock \showarticletitle{Automated test case repair using language models}.
\newblock \bibinfo{journal}{\emph{IEEE Transactions on Software Engineering}} (\bibinfo{year}{2025}).
\newblock


\bibitem[Zhang et~al\mbox{.}(2023b)]%
        {zhang2023algo}
\bibfield{author}{\bibinfo{person}{Kexun Zhang}, \bibinfo{person}{Danqing Wang}, \bibinfo{person}{Jingtao Xia}, \bibinfo{person}{William~Yang Wang}, {and} \bibinfo{person}{Lei Li}.} \bibinfo{year}{2023}\natexlab{b}.
\newblock \showarticletitle{Algo: Synthesizing algorithmic programs with generated oracle verifiers}.
\newblock \bibinfo{journal}{\emph{Advances in Neural Information Processing Systems}}  \bibinfo{volume}{36} (\bibinfo{year}{2023}), \bibinfo{pages}{54769--54784}.
\newblock


\bibitem[Zhang et~al\mbox{.}(2024)]%
        {zhang2024benchmarking}
\bibfield{author}{\bibinfo{person}{Tianyi Zhang}, \bibinfo{person}{Faisal Ladhak}, \bibinfo{person}{Esin Durmus}, \bibinfo{person}{Percy Liang}, \bibinfo{person}{Kathleen McKeown}, {and} \bibinfo{person}{Tatsunori~B Hashimoto}.} \bibinfo{year}{2024}\natexlab{}.
\newblock \showarticletitle{Benchmarking large language models for news summarization}.
\newblock \bibinfo{journal}{\emph{Transactions of the Association for Computational Linguistics}}  \bibinfo{volume}{12} (\bibinfo{year}{2024}), \bibinfo{pages}{39--57}.
\newblock


\bibitem[Zhang et~al\mbox{.}(2023a)]%
        {zhang2023well}
\bibfield{author}{\bibinfo{person}{Ying Zhang}, \bibinfo{person}{Wenjia Song}, \bibinfo{person}{Zhengjie Ji}, \bibinfo{person}{Na Meng}, {et~al\mbox{.}}} \bibinfo{year}{2023}\natexlab{a}.
\newblock \showarticletitle{How well does LLM generate security tests?}
\newblock \bibinfo{journal}{\emph{arXiv preprint arXiv:2310.00710}} (\bibinfo{year}{2023}).
\newblock


\end{thebibliography}
\end{document}